# Spin decomposition and topological properties of spin angular momenta in general electromagnetic fields

Peng Shi, Luping Du*, Aiping Yang, Xiaojin Yin, Xinrui Lei, Xiaocong Yuan*

Nanophotonics Research Centre, Shenzhen Key Laboratory of Micro-Scale Optical Information Technology, Institute of Micro/Nano Optoelectronics, Shenzhen University, 518060, China

*Authors to whom correspondence should be addressed: lpdu@szu.edu.cn, and xcyuan@szu.edu.cn

**Abstract:** Spin angular momenta play important roles in light–matter interactions, which lead to the emergence of spin optics and topological photonics in modern physics. Previously, the spins of plane electromagnetic waves were decomposed into longitudinal and transverse spins owing to its vector characteristics with respect to canonical momentum. Here, we developed a general field theory to reveal the physical origins and topological properties of longitudinal and transverse spins for the arbitrary electromagnetic waves, in whether near-field or free space, and even diversified classical wave fields. Through the decomposition, for the electromagnetic field that carries helicity intrinsically, an extraordinary helicity-dependent transverse spin possessing helicity-dependent spin-momentum locking was discovered. This spin is closely related to the Berry phase and ensures the number of spin-momentum locking states is consistent with the nontrivial spin Chern number. To validate the spin properties, an experiment designed to measure the three-dimensional spin angular momentum densities in an optical focusing configuration was implemented. The findings are important for constructing spin-based field theory and applications exploiting chiral manipulations.

**Introduction:** Spin angular momentum (SAM), a fundamental dynamical property of elementary particles and classical wave fields, plays a critical role in understanding and predicting the behaviors arising from wave–matter interactions [1-7]. In the classical electromagnetic (EM) fields, the SAM associated with the circular polarization orientates in arbitrary directions. In that respect, the widely accepted wisdom was that, for plane-wave solutions of Maxwell's equations, the SAM component oriented along the propagating direction (mean wavevector $\hat{\mathbf{k}}$) is considered to be a longitudinal spin (L-spin) [2]. This L-spin solely depends on the helicity $\sigma$ of circular polarization, which is defined by the EM-field 'self-rotation' within the plane perpendicular to the wavevector [2], whereas the SAM components oriented perpendicular to the mean wavevector represents the helicity-independent transverse spins (T-spin) [8]. To date, T-spins have been investigated in various EM fields, including Gaussian focused fields [9,10], interference fields [11], evanescent waves [12-15], guided modes [16,17], unpolarized fields [18], and photonic chiral spin textures [19-24]. The SAM interact intensively with orbital angular momentum (OAM), especially in the subwavelength scale, raising strong research interest in the spin–orbit interaction (SOI) [2,25] and other remarkable phenomena, offering potential applications in the fields of angular-momentum-based optical manipulation [26,27], unidirectional guided wave [28-31], imaging [32-34], detection and nanometrology [35], and on-chip quantum technologies [36].

However, if complicated structural properties [37], including the inhomogeneities of intensity, phase,

polarization and helicity, are introduced into the EM fields, distinguishing between L-spin and T-spin in the empirical wavevector approach (i.e., longitudinal/transverse meaning parallel/perpendicular to the mean wavevector) brings physical challenges. From the physical point of view, a class of physical quantities should possess a unified physical mechanism embodied by a single universal equation. In quantum physics, photons are spin-1 bosons [2], and hence it is reasonable to extract the L-spin correspondence using concepts in quantum physics; however, for a generic EM field, the physical origin of T-spins and their properties await uncovering quantitatively. For other diverse classical wave fields, such as the acoustic and gravity water waves, the mediating bosons are spin-0 phonons and hence should not possess L-spin. However, both wave fields definitely carry SAM [3-7]. These helicity-independent SAMs correspond to the T-spin of a helicity-independent single polarized EM field and have the same physical origin and unified physical properties.

Here, we construct a unified field theory based on the decomposition of the SAM for a generic EM field into L-spin and T-spin, that enable their physical origins and accompanying intrinsic topological properties to be uncovered. The decomposition technique is similar to that for the Helmholtz decomposition of an arbitrary vector field, and therefore the theory can be applied to SAM decompositions for acoustic and gravity water waves. The equations reveal that L-spin is associated with the helicity that is oriented parallel to the local wavevector or the Minkowski canonical momentum, whereas T-spin stems from the inhomogeneity of the momentum density of the field and locks with the kinetic Abraham momentum in the near field or free space. Remarkably, if an inhomogeneity of the helicity-related momentum density is present in a structured EM field, an extraordinary helicity-dependent T-spin appears and leads simultaneously to helicity-dependent spin-momentum locking. Under this circumstance, the number of spin-momentum locking states is consistent with the nontrivial topological spin Chern number of the EM field. In addition, the helicity-dependent T-spin refers to the inverted helical component in the EM system and therefore is closely related to the evolution of the geometric phase in optical systems. More curiously, this decomposition of the spin vector results in a kind of T-spin oriented parallel to the mean wavevector and L-spin oriented perpendicular to the mean wavevector, which demonstrate the empirical wavevector approach definitely faces challenges in separating the L-spin and T-spin of structured light fields. We furthermore demonstrate the spin properties and the accompanied spin-momentum locking properties experimentally by mapping the three components of SAM in an optical focused beam with circular polarizations in our in-house developed near-field imaging system. Our findings deepen the understanding of the underlying physics of spins for all classical wave fields and open an avenue for applications including manipulation and robust chiral devices.

**Results:** For a general structured EM wave carrying helicity and inhomogeneity simultaneously, we proved theoretically that the T-spin ($\mathbf{S}_t$) and L-spin ($\mathbf{S}_l$) of an arbitrary EM wave can be determined by (Supplementary Sections 1-5)

$$\mathbf{S}_l = \sum_i \hbar \sigma_i \hat{\mathbf{k}}_i + \sum_{i \neq j} \hbar \sigma_{ij} \hat{\mathbf{k}}_{ij} , \tag{1}$$

and

$$\mathbf{S}_t = \frac{1}{2k^2} \nabla \times \mathbf{\Pi} . \tag{2}$$

Here, we consider a generic EM field for which the electric and magnetic field components can be expanded into superpositions of a plane-wave basis, and carry helicity $\sigma_i$ and local wavevector $\hat{\mathbf{k}}_i$ for each elementary plane wave $i$, and helicity $\sigma_{ij}$, and local wavevector $\hat{\mathbf{k}}_{ij}$ for the coupling of interfering plane waves $i$ and $j$.

We must emphasize that the interfering terms appear for the non-orthogonal plane-wave basises. The total spin is thus given by $\mathbf{S} = \mathbf{S}_l + \mathbf{S}_t = \langle\psi|\bar{\mathbf{S}}|\psi\rangle/\hbar\omega$ and $\bar{\mathbf{S}} = [\hat{\mathbf{S}}, \mathbf{0}; \mathbf{0}, \hat{\mathbf{S}}]$ with $\hat{\mathbf{S}}$ representing the spin-1 matrix in SO(3) [38], **0** the 3×3 zero matrix, and $|\psi\rangle$ the classical photon wave function analogous to the quantum wave function [38].

To detailly analyse the physical properties of L-spin in Eq. (1), we first take an elliptically polarized plane wave propagating in the *x*-direction with the electric and magnetic fields

$$\mathbf{E}(\mathbf{r}) = \left(0\hat{\mathbf{x}} + A_s\hat{\mathbf{y}} + A_p\hat{\mathbf{z}}\right)e^{i(kx-\omega t)}$$
$$\mathbf{H}(\mathbf{r}) = \left(0\hat{\mathbf{x}} - A_p\hat{\mathbf{y}} + A_s\hat{\mathbf{z}}\right)e^{i(kx-\omega t)}/\eta \qquad (3)$$

defined in an arbitrary orthogonal coordinate (*x*, *y*, *z*) (Supplementary Section 2) for example. In quantum physics, photons are spin-1 bosons with the direction of spin parallel to that of the photon momentum [38]. Here, the total SAM of this plane wave is $\mathbf{S} = \hbar\sigma\hat{\mathbf{k}}$ where $\sigma = Im\{A_s^*A_p - A_p^*A_s\}/\{A_s^*A_s + A_p^*A_p\}$ is the helicity of a single wave-packet (also termed the polarization ellipticity [27]). Here, $A_s$ and $A_p$ denote the amplitudes of the *s*-component and *p*-component, $\eta$ the wave impedance, $\hbar$ reduced Planck constant, $\omega$ the angular frequency, and $k$ the wavenumber. The special instances $\sigma = \pm 1$ represent the two circularly polarized (CP) modes of light corresponding to the two helical states in quantum physics [2]. Thus, the expression for the SAM helps delineate the global properties of EM fields from the perspective of classical field theory as well as the elementary dynamical properties of optical wave-packets from the viewpoint of quantum theory. The spin vector of CP light is parallel to the local wavevector $\hat{\mathbf{k}}$ and thus was regarded previously as L-spin. In physics, for theoretical consistency, the elementary feature of L-spin in a generic EM field should coincide with the definition of the photonic spin in quantum physics, i.e., the L-spin is parallel to the local wavevector $\hat{\mathbf{k}}$ and its intensity is determined by the helicity $\sigma$ of the EM field.

However, it is worth noting that L-spin is based on the link between the EM helicities for each of the elementary interfering waves and their local wavevectors $\hat{\mathbf{k}}$, rather than the mean wavevector. In our work, the mean wavevector is given by the canonical momentum of the total EM field whereas the local wavevector is identified by the canonical momentum of each plane-wave component when expanding the EM field into a superposition of plane waves. Thus, the first term on the right side of equation (1) represents the summation of L-spins of each elementary plane wave and the second term represents the sum of L-spins from couplings between the interfering plane waves. The appearance of the coupling term stems from the non-orthogonality of the coupled plane-wave basis, and therefore it can vanish if the two plane-wave basises are orthogonal (such as left- and right-handed circular polarizations or the examples considered in **Fig. 3**).

On the other hand, to uncover the physical origin of T-spin, we consider a classical hydrodynamic model in which a particle is immersed in a fluid possessing a gradient field of momentum [**Fig. 1**]. Assuming the flow of the water wave is in the +*x*-direction with its momentum density decreasing in the *y*-direction, the immersed particle experiences an anticlockwise transverse torque (**M**$_z$), its intensity being proportional to the local gradient of the momentum density in the *y*-direction [**Fig. 1(a)**]. If the flow is in the +*y*-direction with momentum density increasing in the *x*-direction, the immersed particle also experiences an anticlockwise transverse torque (**M**$_z$) with an intensity proportional to the local gradient of momentum density in the *x*-direction [**Fig. 1(b)**]. In total, the particle immersed in the fluid flow with a gradient momentum density experiences a transverse torque, the intensity of which is proportional to the vorticity associated with the momentum density. Correspondingly, we conjecture that the generation of T-spin in an EM system is related to the vorticity associated with the momentum flow of the photons.

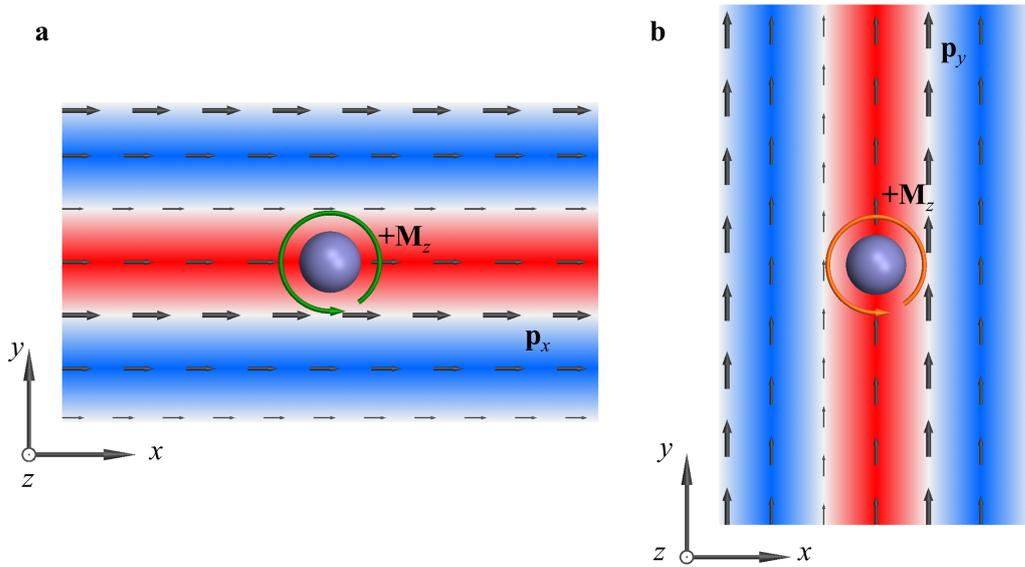

**Fig. 1. Hydrodynamic model used to reveal the relationship between transverse torque M$_z$ and momentum flow:** (a) momentum flow propagating along the +*x*-direction with magnitude gradually decreasing in the *y*-direction causes the immersed particle to rotate anticlockwise; (b) momentum flow propagating along the +*y*-direction with magnitude gradually increasing in the *x*-direction also causes the same particle to rotate anticlockwise. The overall spinning effect on the immersed particle is therefore related to the vorticity of momentum flow. This theoretical analysis is consistent with the spin-momentum relation of deep-water gravity wave: $\mathbf{S}_{GW} = \nabla_2 \times \mathbf{\Pi}_{GW}/2k_{GW}^2$. The magnitude of each arrow indicates the intensity of energy flow and the direction of energy flow is given by the arrow's orientation. The background colour indicates the *z*-component SAM density, with red and blue indicating the positive and negative SAM densities, respectively.

In classical field theory, there are three types of momentum densities [2,6,25]: the kinetic momentum **Π** of photons or phonons associated with the energy flow of wave fields, which can be decomposed into the canonical momentum **P** and the Belinfante spin momentum **P**$_s$. These three momenta describing the flow of photons or phonons are our candidates to evaluate T-spin. Previously, for plane-waves of an EM system, only the canonical momentum **P** associated with the local wavevector was employed to identify the T-spin in various ways [27]. However, for these diverse classical wave fields, basic physical challenges are faced when searching for a unified physical mechanism to evaluate T-spin in a universal manner because spin momentum appears that also plays a critical role in T-spin generation. For example, for plane-waves of an EM field that has pure L-spin (**S** ∥ **k̂**), one easily derives that the spin momentum **P**$_s$ = ∇ × **S**/2 is perpendicular to the wavevector **k̂** (**P**$_s$ ⊥ **k̂**). This is illogical because spin momentum should also be longitudinal; thus, **P**$_s$ vanishes for the field with pure L-spin. In contrast, for an inhomogeneous structured EM field containing T-spins, the spin momentum **P**$_s$ = ∇ × **S**/2 is longitudinal, demonstrating that spin momentum is prevalent in a structured field and, inevitably connected with T-spin. In particular, for a relativistic field such as an EM wave, the canonical group velocity determined by the canonical momentum **P** would be superluminal [39]. This contradicts a principle of relativity, and hence a spin momentum should appear and be antiparallel to the canonical momentum to guarantee the total group velocity associated with the kinetic momentum is subluminal [39]. In other words, spin momentum is highly related to the T-spin in a structured field and therefore kinetic momentum **Π** = **P** + **P**$_s$, which combines canonical and spin momenta, is reasonable when employed to evaluate the T-spin in general scenarios.

To verify the above analysis, we proved theoretically that the spin-momentum property for the two-dimensional motion of surface water waves [6] in the *xy*-plane [**Fig. 1**, z = 0] obeys the relation,

$$\mathbf{S}_G = \frac{1}{2k_G^2}\nabla_2 \times \mathbf{\Pi}_G \tag{4}$$

where $\mathbf{S}_G$ and $\mathbf{\Pi}_G$ are the total SAM and kinetic momentum, respectively, of monochromatic time-harmonic gravity water waves. Here, $\nabla_2 = (\partial_x, \partial_y)$; $\omega_G^2 = gk_G$ with $\omega_G$ and $k_G$ the angular frequency and wavenumber of the water wave, and *g* denotes the gravitational acceleration. Equation (4), which verifies our analysis accurately, demonstrates that the SAM of a surface water wave is locked with the kinetic momentum and obeys the right-hand rule; the total SAM may be considered as the T-spin. This conclusion is consistent with the intrinsic helical property of the corresponding phonons; the surface deep-water gravity wave are spin-0 phonons ($\sigma_G = 0$) and do not possess L-spin. A spin-momentum relation is also obtained for longitudinal acoustic waves, specifically, $\mathbf{S}_A = \nabla \times \mathbf{\Pi}_A / k_A^2$, which also indicates the transversality of SAM for a spin-0 longitudinal phonon ($\sigma_A = 0$) (**Table 1**). Here, $\mathbf{S}_A$ and $\mathbf{\Pi}_A$ are, respectively, the total SAM and kinetic momentum of a monochromatic time-harmonic acoustic wave.

**Table 1.** The dynamical and topological properties of generic EM wave, single polarized evanescent EM wave, deep-water gravity wave, and acoustic wave fields.

| | Generic EM wave | Single polarized evanescent EM wave | Gravity water wave | Acoustic wave |
|---|---|---|---|---|
| Field components | Electric field **E**; Magnetic field **H**; | Electric or magnetic Hertz potential $\Psi$; | In-plane velocity **V**; Normal velocity *W*; | Velocity **v**; Pressure *p*; |
| Kinetic momentum | $\mathbf{\Pi} = \frac{1}{2c^2}\text{Re}\{\mathbf{E}^* \times \mathbf{H}\}$ | $\mathbf{\Pi} = \frac{\varepsilon k^2 k_p^2}{2\omega}\text{Im}\{\Psi^*\nabla\Psi\}$ | $\mathbf{\Pi}_G = \frac{\rho_G k_G}{\omega_G}\text{Im}\{W^*\mathbf{V}\}$ | $\mathbf{\Pi}_A = \frac{1}{2c_A^2}\text{Re}\{p^*\mathbf{v}\}$ |
| Spin angular momentum | $\mathbf{S} = \frac{1}{4\omega}\text{Im}\left\{\begin{array}{l}\varepsilon\mathbf{E}^* \times \mathbf{E} \\ +\mu\mathbf{H}^* \times \mathbf{H}\end{array}\right\}$ | $\mathbf{S} = \frac{\varepsilon k_p^2}{4\omega}\text{Im}\{\nabla\Psi^* \times \nabla\Psi\}$ | $\mathbf{S}_G = \frac{\rho_G}{2\omega_G}\text{Im}\{\mathbf{V}^* \times \mathbf{V}\}$ | $\mathbf{S}_A = \frac{\rho_A}{2\omega_A}\text{Im}\{\mathbf{v}^* \times \mathbf{v}\}$ |
| Helicity | Spin-1 photon $\sigma = \pm 1$ | Spin-1 photon $\sigma = 0$ | Spin-0 phonon $\sigma_G = 0$ | Spin-0 phonon $\sigma_A = 0$ |
| Spin-momentum locking | $\mathbf{S}_t = \frac{1}{2k^2}\nabla \times \mathbf{\Pi}$ $\mathbf{S}_l = \sum_i \hbar\sigma_i\hat{\mathbf{k}}_i + \sum_{i \neq j}\hbar\sigma_{ij}\hat{\mathbf{k}}_{ij}$ | $\mathbf{S}_t = \frac{1}{2k^2}\nabla \times \mathbf{\Pi}$ $\mathbf{S}_l = 0$ | $\mathbf{S}_G = \frac{1}{2k_G^2}\nabla_2 \times \mathbf{\Pi}_G$ | $\mathbf{S}_A = \frac{1}{k_A^2}\nabla \times \mathbf{\Pi}_A$ |

The kinetic momentum and SAM properties of water wave and acoustic wave can be found in Ref. (7) and the Supplemental Table. S1 of Ref. [15]. The kinetic momentum and SAM properties of single polarized evanescent EM wave can be found in Ref. [17]. For the longitudinal acoustic wave, $c_A^2 = 1/\beta_A\rho_A$ is the speed of acoustic wave with $\beta_A$ the compressibility of acoustic medium; $\rho_A$ is the mass density of the acoustic medium; $\omega_A$ and $k_A = \omega_A/c_A$ are the angular frequency and wavenumber; $\sigma_W = 0$ for the phonons corresponding to the longitudinal acoustic waves. $\rho_G$ is the mass density of the fluid.

Thus, Eq. (2) reveals that, for an EM wave, the T-spin arises from the inhomogeneous momentum flow density of the EM field and its transversality constraint satisfies owing to the identity $\nabla \cdot (\nabla \times \mathbf{A}) = 0$ for an arbitrary vector field **A**. Moreover, the T-spin is locked to the kinetic momentum [13-17] in a manner unrelated to the L-spin of the EM wave propagating in a homogeneous medium. This spin-momentum locking originates from the intrinsic spin–orbit coupling in Maxwell's equations and is considered as a fundamental property of T-spin for an arbitrary EM field, either propagating in free space or confined at an interface (evanescent waves). More interestingly, in consequence of the long-standing Abraham–Minkowski debate [16,40], the accepted wisdom is that the Minkowski-type canonical momentum determines the local wavevector of photons and the Abraham kinetic momentum is always associated with the group velocity (called a current by M.V. Berry [38]) of EM wave fields. Therefore, spin-momentum locking between the kinetic momentum and SAM can also be regarded as spin-current locking and thus is different from the quantum spin-Hall effect in condensed matter physics [41].

This result is also consistent with the conclusion of the Abraham–Minkowski debate [40]: the Minkowski canonical momentum determines the local wavevector of photons and is reasonable to evaluate the L-spin. Remarkably, because the local wavevector $\hat{\mathbf{k}}$ is proportional to the canonical momentum given by the inner product of momentum operator $\hat{\mathbf{P}} = -i\hbar\nabla$, and considering the identity $\nabla \times \nabla\psi = 0$ for an arbitrary scalar field $\psi$, the decomposition of the spin vector into L-spin and T-spin is similar to the Helmholtz decomposition of an arbitrary vector [42], thereby making the notion of transversality and longitudinality consistent with other physical systems. This consistency is the fundamental starting point of our research, and we summarize the spin-momentum properties of generic EM waves, single polarized evanescent waves (which can be considered as helicity-independent), water waves, and acoustic waves (see **Table 1** for a comparison).

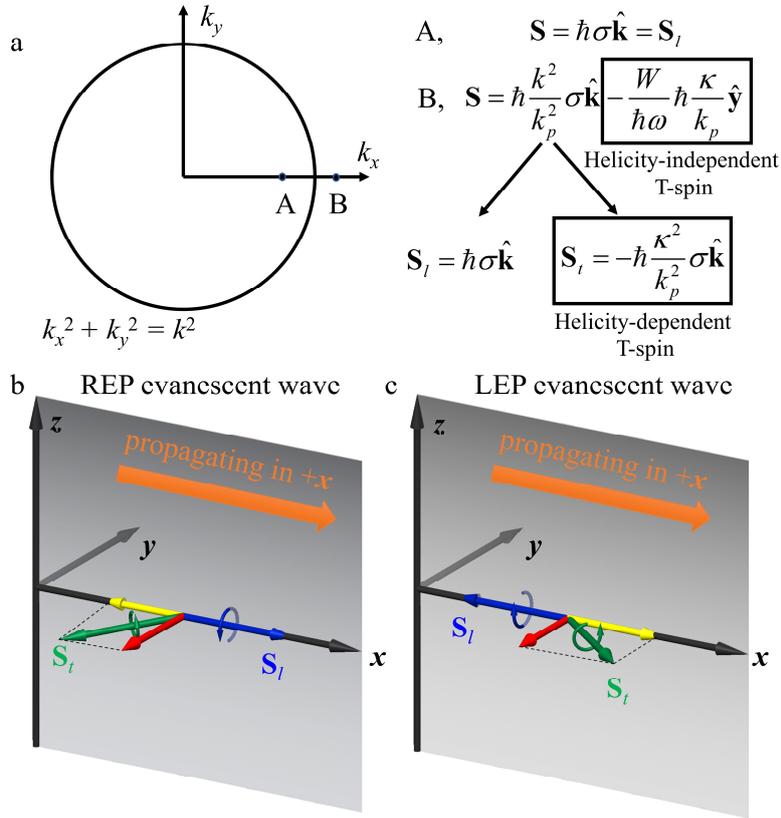

**Fig. 2. Spin decomposition of an elliptically-polarized plane wave.** (a) k-space representation of the plane waves, in which the wavenumbers confined within the circle $k_x^2 + k_y^2 = k^2$ correspond to waves propagating in free space whereas those outside of the circle correspond to evanescent waves confined to an interface. The consistency of the elementary feature of L-spin for the same kind of wave-packets (points "A" and "B") leads to the classification of T-spin into the normal helicity-independent component aligned perpendicular to the wavevector (the y-component) and the extraordinary helicity-dependent component aligned anti-parallel to the wavevector. (b-c), Illustrations of the generic spin properties of a right-handed and left-handed elliptically polarized evanescent plane-wave, respectively, propagating along the +x-direction. If the kinetic momentum is reversed, the T-spins (green arrows), including the helicity-independent T-spin (red arrows) and the helicity-dependent T-spin (yellow arrows), become opposite. Therefore, four spin-momentum locking states exist in a general EM system, consistent with the $\mathbb{Z}_4$ topological invariance of the optical wave-packet [13,17].

To further understand the decomposition of spins of complex EM fields into L-spin and T-spin using our theory, and more importantly the helicity-induced extraordinary properties when compared with spin-0 phonons (water and acoustic waves), we first take the elliptically polarized plane wave as an example. A plane wave either propagates or evanesces depending on the wavevector components [**Fig. 2(a)**]. Assuming that the wave is propagating along the *x*-direction, for a plane wave in free space as demonstrated in equation (1) [point "A" in **Fig. 2(a)**], the SAM of the wave is $\mathbf{S} = \hbar\sigma\hat{\mathbf{k}}$ and is pure L-spin. Now consider its evanescent counterpart [point "B" in **Fig. 2(a)**] with electric and magnetic fields

$$\mathbf{E}(\mathbf{r}) = \left(-A_p i\kappa/k\,\hat{\mathbf{x}} + A_s\hat{\mathbf{y}} + A_p k_p/k\,\hat{\mathbf{z}}\right)e^{(ik_p x - \kappa z)}$$
$$\mathbf{H}(\mathbf{r}) = \left(-A_s i\kappa/k\,\hat{\mathbf{x}} - A_p\hat{\mathbf{y}} + A_s k_p/k\,\hat{\mathbf{z}}\right)e^{(ik_p x - \kappa z)}/\eta$$
(5)

The SAM is found to be $\mathbf{S} = \hbar(k^2/k_p^2)\sigma\hat{\mathbf{k}} - W\kappa/\omega k_p\,\hat{\mathbf{y}}$, where $k_p$ denotes the horizontal wavenumber, $i\kappa$ represents the wavenumber in the *z*-direction with $k_p^2 = k^2 + \kappa^2$, and $W$ denotes the time-averaged energy density. We observe that the *y*-component of the SAM corresponds to the normal helicity-independent T-spin of a single polarized evanescent plane wave (transverse magnetic (TM) or transverse electric (TE) polarization) [17] and is perpendicular to the local wavevector $\hat{\mathbf{k}}$. However, compared with the EM helicity of a propagating elliptically polarized plane wave, the SAM component parallel to the local wavevector contains an additional factor: $k^2/k_p^2$, which is illogical in physics if we consider it entirely as a L-spin because the elementary feature of L-spin for the same kind of wave-packet should be constant [**Fig. 2(a)**]. Indeed, given the evanescent property of the wave in the *z*-direction, one finds two types of kinetic momentum density (Supplementary Section 3, Eq. (S26)): the helicity-unrelated *x*-component of momentum and the helicity-related *y*-component of the momentum density, both of which decay in the *z*-direction. Thus, one can expect two components of T-spin,

$$\mathbf{S}_t = \frac{1}{2k^2}\nabla\times\mathbf{\Pi} = -\hbar\frac{\kappa^2}{k_p^2}\sigma\hat{\mathbf{k}} - \frac{W}{\omega}\frac{\kappa}{k_p}\hat{\mathbf{y}},$$
(6)

whereas the L-spin is $\mathbf{S}_l = \mathbf{S} - \mathbf{S}_t = \hbar\sigma\hat{\mathbf{k}}$, which now coincides with that in free space. Interestingly, asides from the helicity-independent T-spin that was investigated intensively in the past [8-18], an extraordinary hidden T-spin that is helicity-dependent is predicted from the theory. This leads to four spin-momentum locking states for a generic EM field [**Figs. 2(b, c)**] for which the respective spin properties of the right-handed and left-handed elliptically polarized states propagating in the +*x*-direction are shown. If the kinetic momentum associated with the flow of photons is reversed, both the helicity-dependent and helicity-independent T-spins are inverted simultaneously. This indicates that the general EM field possesses $\mathbb{Z}4$ topological invariance, which is consistent with a nontrivial spin Chern number of light [17]. When the dual symmetry between the electric and magnetic constitutive relations is broken [13,43], and only a single polarized state survives, the four spin-momentum locking states downgrade to two helicity-independent states, as indicated in the third column of **Table 1**.

The aforementioned concepts of EM spin can be generalized to an arbitrary EM wave field by expanding it into the superposition of plane waves, for either near field or free space. Here, for simplicity, we only consider evanescent plane waves as an example and demonstrate two-wave interference [**Figs. 3(a, f, k)**]. The results can be extended to multiple wave interference and thus an arbitrary EM field. Assume the two interfering fields are

$$\mathbf{E}_1(\mathbf{r}) = \left(-A_{p1} i\kappa/k\,\hat{\mathbf{x}} + A_{s1}\hat{\mathbf{y}} + A_{p1} k_p/k\,\hat{\mathbf{z}}\right)e^{(ik_p x - \kappa z)}$$
$$\mathbf{E}_2(\mathbf{r}) = \left(-A_{p2} i\kappa/k\,\hat{\mathbf{x}} + A_{s2}\hat{\mathbf{y}} + A_{p2} k_p/k\,\hat{\mathbf{z}}\right)e^{(ik_p x - \kappa z)}$$
(7)

and are rotated respectively through angle $+\theta$ and $-\theta$ with respect to the $x$ axis. Then, the total electric field of the superposed field is expressible as (Supplementary Section 4)

$$\mathbf{E} = \hat{R}_z(-\theta)\mathbf{E}_1\left[\hat{R}_z(\theta)\mathbf{r}\right] + \hat{R}_z(\theta)\mathbf{E}_2\left[\hat{R}_z(-\theta)\mathbf{r}\right], \tag{8}$$

Here, $\hat{R}_z(\theta)$ denotes the rotational operator with respect to the $z$ axis; $r = (x,y,z)$ represents the coordinates. The magnetic field is calculated using Faraday's law of electromagnetic induction, $\mathbf{H} = \nabla \times \mathbf{E}/i\omega\mu$. One then obtains an energy density containing three parts $W = W_1 + W_2 + W_c$, for which

$$W_1 = \frac{\varepsilon k_p^2}{2k^2}\left\{A_{p1}^*A_{p1} + A_{s1}^*A_{s1}\right\}e^{-2\kappa z} \tag{9}$$

and

$$W_2 = \frac{\varepsilon k_p^2}{2k^2}\left\{A_{p2}^*A_{p2} + A_{s2}^*A_{s2}\right\}e^{-2\kappa z} \tag{10}$$

denote the energy densities of waves 1 and 2, and

$$W_c = \frac{\varepsilon k_p^2}{2k^2}\left\{\left(A_{p1}^*A_{p2}e^{-2ik_{s1}y} + A_{p2}^*A_{p1}e^{+2ik_{s1}y}\right) + \left(A_{s1}^*A_{s2}e^{-2ik_{s1}y} + A_{s2}^*A_{s1}e^{+2ik_{s1}y}\right)\right\}e^{-2\kappa z} \tag{11}$$

denotes the coupling energy density with $k_{s1} = k_p\sin\theta$ and $k_{p1} = k_p\cos\theta$. Note that $W_c$ is local and its integral over the whole $xy$-plane vanishes. Based on this decomposition of the energy density, the unit directional vector of the mean wavevector of the superposed field is also decomposable to $\hat{\mathbf{k}} = \hat{\mathbf{k}}_1 + \hat{\mathbf{k}}_2 + \hat{\mathbf{k}}_c$, for which the local wavevectors of waves 1 and 2 are

$$\hat{\mathbf{k}}_1 = \frac{W_1}{\hbar\omega}\left(\frac{k_{p1}}{k_p}\hat{\mathbf{x}} + \frac{k_{s1}}{k_p}\hat{\mathbf{y}} + 0\hat{\mathbf{z}}\right) \tag{12}$$

and

$$\hat{\mathbf{k}}_2 = \frac{W_2}{\hbar\omega}\left(\frac{k_{p1}}{k_p}\hat{\mathbf{x}} - \frac{k_{s1}}{k_p}\hat{\mathbf{y}} + 0\hat{\mathbf{z}}\right), \tag{13}$$

and

$$\hat{\mathbf{k}}_c = \frac{W_c}{\hbar\omega}\left(\frac{k_{p1}}{k_p}\hat{\mathbf{x}} + 0\hat{\mathbf{y}} + 0\hat{\mathbf{z}}\right) \tag{14}$$

represents the local wavevector of the coupling energy density by comparing the total energy density and mean wavevector of the superposed field. In this way, the L-spin can be calculated as: $\mathbf{S}_l = \hbar\sigma_1\hat{\mathbf{k}}_1 + \hbar\sigma_2\hat{\mathbf{k}}_2 + \hbar\sigma_c\hat{\mathbf{k}}_c$, where the three helicities are

$$\sigma_1 = \frac{\text{Im}\left\{A_{s1}^*A_{p1} - A_{s1}A_{p1}^*\right\}}{A_{p1}^*A_{p1} + A_{s1}^*A_{s1}}, \tag{15}$$

$$\sigma_2 = \frac{\text{Im}\left\{A_{s2}^*A_{p2} - A_{s2}A_{p2}^*\right\}}{A_{p2}^*A_{p2} + A_{s2}^*A_{s2}}, \tag{16}$$

and

$$\sigma_c = \frac{\text{Im}\left\{\left(A_{s1}^*A_{p2}e^{-2ik_{s1}y} - A_{s1}A_{p2}^*e^{+2ik_{s1}y}\right) + \left(A_{s2}^*A_{p1}e^{+2ik_{s1}y} - A_{s2}A_{p1}^*e^{-2ik_{s1}y}\right)\right\}}{\left(A_{p1}^*A_{p2}e^{-2ik_{s1}y} + A_{p2}^*A_{p1}e^{+2ik_{s1}y}\right) + \left(A_{s1}^*A_{s2}e^{-2ik_{s1}y} + A_{s2}^*A_{s1}e^{+2ik_{s1}y}\right)}, \tag{17}$$

respectively. The helicity of each individual wave is given by the corresponding polarization ellipticity [27].

Thus, the link between the EM helicities and their local wavevectors is intrinsically based on the decomposition of the energy density and mean wavevector. The same conclusion can be reached for waves in free space by calculating two-wave interference (Supplementary Section 5).

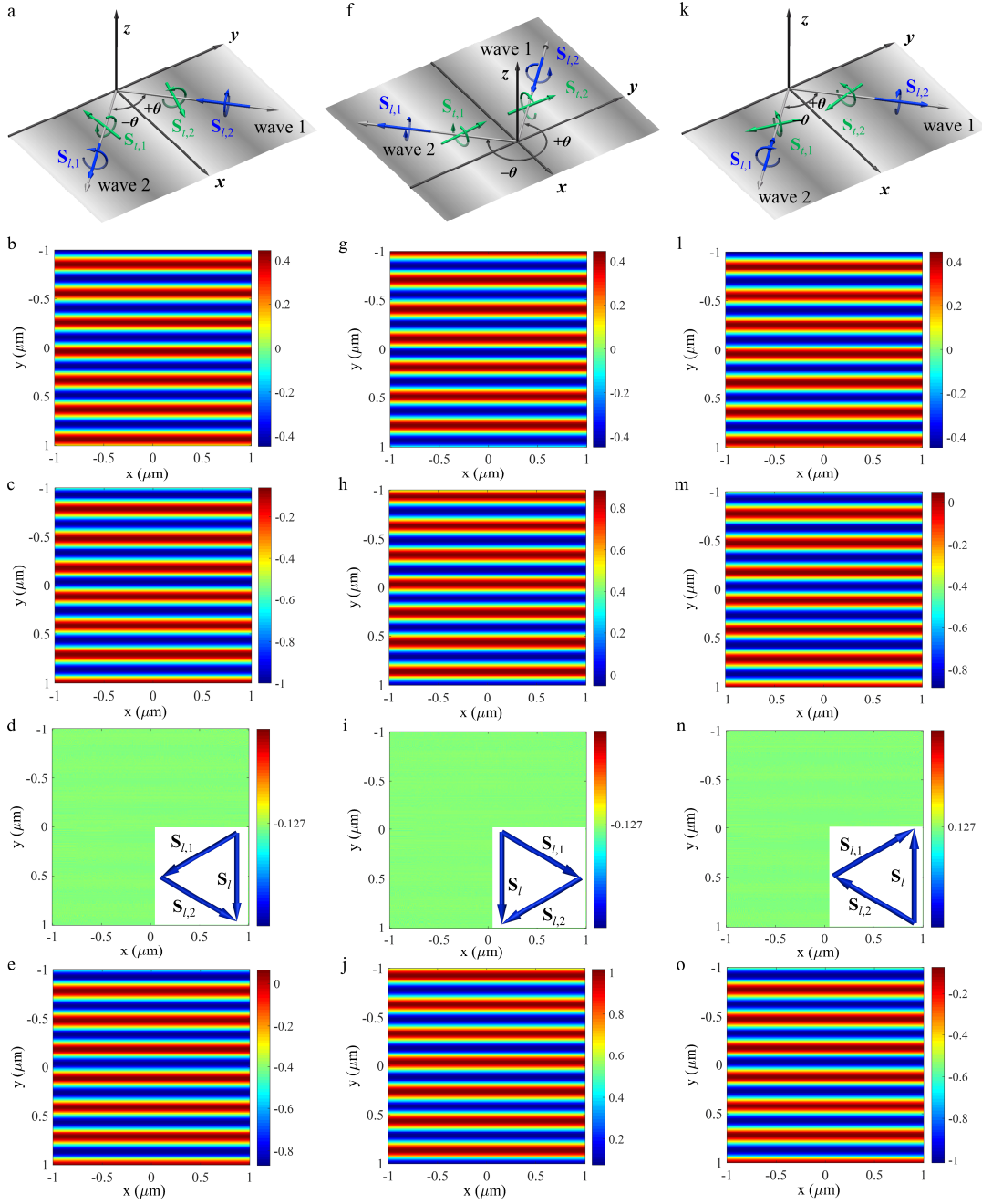

**Fig. 3. Spin properties of interference fields between two elliptically polarized evanescent plane waves.** (a) Schematic of the interference between two evanescent plane waves carrying opposite helices in the *xy*-plane. (b-c) Spatial distributions of the *z*- and *y*-component of the SAMs when $\theta = 45°$, $A_{s1} = 5+2i$ and $A_{s2} = 5–2i$. (d) Extracted L-spin from the *y*-component of the SAM; inset shows the vector decomposition, and (e) the remaining T-spin for the *y*-component of the SAM. (f-j) same as (a-e), but with $\theta = 135°$. (k-o) same as (a-e), but with the opposite helicities, i.e., $A_{s1} = 5–2i$ and $A_{s2} = 5+2i$. In the calculation, $A_{p1} = A_{p2} = 1$, $k_p = 1.5k$ and the wavelength of the waves is 632.8nm.

To understand in detail the spin property of an EM field, we first consider the interference of two waves with opposite helicities [**Fig. 3(a)**]. In this way, the coupling helical term σc vanishes and thereby beneficial when analysing the spin property of EM fields. Assuming $A_{p1} = A_{p2} = 1$, $A_{s1} = 5+2i$, $A_{s2} = 5-2i$, and the propagating angles $\theta$ of the two plane waves are +45º and –45º, the canonical momentum associated with the mean wavevector is along the +x-direction and varies periodically in the y-direction [Supplementary Section 4, **Fig. S5(e)**]. The kinetic momentum has two components: the helicity-unrelated component along the +$x$-direction and varying periodically in the y-direction [Supplementary Section 4, **Fig. S5(c)**], and the helicity-related component along the –$x$-direction and being homogeneous in the $xy$-plane [Supplementary Section 4, **Fig. S5(d)**]. All the momenta decay exponentially in the $z$-direction. In this instance, the $x$-component SAM is absent, and only the $z$- and $y$-components of SAM arise [**Figs. 3(b, c)**, respectively]. From our theory, the $z$-component SAM is a pure helicity-independent T-spin because the helicity-related kinetic momentum is spatially invariant in the $xy$-plane, whereas the $y$-component SAM contains both L-spin and T-spin [**Figs. 3(d, e)**, respectively].

To verify this, we then consider the interference of these two waves by changing the propagating angles to +135º and –135º [**Fig. 3(f)**] so that the propagating direction given by the canonical momentum is opposite to that given in **Fig. 3(a)**. Of note in this instance is that the helicity-unrelated kinetic momentum is inverted [Supplementary Section 4, **Fig. S5(h)**], whereas the helicity-related kinetic momentum remains unchanged [Supplementary Section 4, **Fig. S5(i)**]. Thus, one finds that the z-component of SAM in **Fig. 3(g)** is exactly inverted to that in **Fig. 3(b)**. This is a manifestation of spin-momentum locking of the helicity-independent T-spin. The L-spin present in the y-component of SAM can be determined from a vector analysis based on equation (4) [see insets in **Figs. 3(d, i)**]. The identity of the spins in **Figs. 3(d, i)** demonstrate that L-spin is independent of the propagating direction given by the canonical momentum and does not possess the spin-momentum locking property. The resultant found by subtracting the L-spins from the overall y-component of the SAMs thus yields the properties of the T-spin. Moreover, the variation in colourbar values in **Figs. 3(e, j)** indicates that the resultant T-spins should contain both helicity-independent and helicity-dependent T-spins because a pure helicity-independent T-spin is reversed exactly when the propagation direction is reversed. From equation (4), the helicity-independent y-component of T-spins were generated through the decay of the helicity-unrelated kinetic momentum along the z-direction and are opposite when reversing the propagating direction of the field [Supplementary Section 4, **Figs. S6(b, f)**]. In contrast, the y-component helicity-dependent T-spins are induced by the helicity-related component of kinetic momentum, which although being invariant in the $xy$-plane, decays in the $z$-direction. The helicity-dependent T-spins remain unchanged when changing the propagating direction because the helicity-related kinetic momentum remains unchanged [Supplementary Section 4, **Figs. S6(c, g)**].

Furthermore, to illustrate the helicity-dependent property of the EM spins, we consider the interference of two waves [**Fig. 3(a)**] exchanging their helical properties [**Fig. 3(k)**]. In this instance, the direction of the canonical momentum is similar to that in **Fig. 3(a)** except for a translation in the $y$-direction. Here, we shift the calculation region of the SAMs to eliminate the effect of this translation. Compared with **Fig. 3(a)**, the helicity-unrelated kinetic momentum remains unchanged [Supplementary Section 4, **Figs. S5(c, m)**] whereas the helicity-related kinetic momentum is reversed [Supplementary Section 4, **Figs. S5(d, n)**]. Thus, the helicity-independent T-spin in **Fig. 3(l)** is the same as that in **Fig. 3(b)**, whereas the L-spin in **Fig. 3(n)** from the vector analysis is inverted to that in **Fig. 3(d)**. By subtracting the L-spins from the $y$-component of SAM [**Fig. 3(m)**], one obtains the T-spin in the $y$-direction [**Fig. 3(o)**] as well as the corresponding extracted helicity-independent and helicity-dependent T-spins [Supplementary Section 4, **Figs. S6(j, k)**]. Under this circumstance, the helicity-independent T-spin remains unchanged whereas the helicity-dependent T-spin is reversed. For a clear comparison, we summarize the primary properties of these three types of EM spins in **Table 2**.

Table 2. Physical properties of T-spin and L-spins for a generic EM field

| Classifications | Spin-momentum locking? | Helicity-dependent? |
|---|---|---|
| L-spin | No | Yes |
| Helicity-dependent T-spin | Yes | Yes |
| Helicity-independent T-spin | Yes | No |

The T-spin is closely related to the Berry curvature of an optical system. For a single polarized EM field, the kinetic momentum is expressed as $\mathbf{\Pi} \propto \langle \Psi | i\nabla | \Psi \rangle$ with $|\Psi\rangle$ representing the potential [44]. The T-spin is then given by $\mathbf{S}_t \propto \nabla \times \mathbf{\Pi} \propto \langle \nabla\Psi | \times i | \nabla\Psi \rangle$, which has a similar form to the Berry curvature of the potential [2,35]. Moreover, for a generic EM field, the T-spin determined by $\nabla \times \mathbf{\Pi}$ also has a similar structure as the quantum 2-form [45] that generates the Berry phase associated with a circuit (Supplementary Section 7). For a single polarized EM field, the circulation integral of the Berry curvature defining the geometric phase vanishes, and thus the helicity-independent T-spin is unrelated to the geometric phase. However, from Eq. 5 for a general EM field, the helicity-dependent T-spin is found to be antiparallel to the local wavevector. This is indeed a general property of the helicity-dependent T-spin and widely exists in a generic EM field (see the two-wave interferences in Supplementary Sections 4 and 5). Previously, the generation of this inverted helical component was explained based on the evolution of the geometric phase in EM systems [2]. This demonstrates that the helicity-dependent T-spin is closely related to the evolution of the geometric phase in EM systems such as focused CP beams (Supplementary Section 8). Based on the former considerations, we formulated four Maxwell-like spin-momentum equations and a Helmholtz-like equation in Supplementary Section 6 that can be utilized to analyse the spin-orbit coupling properties for general EM fields.

Finally, to validate the above intriguing properties of EM spins, we built a scanning imaging system to map the three components of SAM for focused CP light propagating in the *z*-direction (Supplementary Sections 9 and 10). **Figures 4(a, b)** and **4(g, h)** exhibit the theoretical calculated results of $S_x$ and $S_y$ for right-handed circular polarization (RCP) and left-handed circular polarization (LCP) focused light, respectively; the corresponding experimental results are shown in Figs. **4(d, e)** and **4(j, k)**. The experimental results match well with the theoretical calculated results and reveal that these two SAM components of a focused field remain unchanged when the incident light is converted from RCP to LCP. These results correspond to the helicity-independent T-spin. **Figures 4(c, i)** exhibits the theoretical calculated results of the z-component SAMs for focused RCP and LCP light, respectively. From the corresponding experimental results [**Figs. 4(f, l)**], we observed that the z-component SAMs are helicity-dependent and inverted when the incident light is converted from RCP to LCP. From the detailed calculations in Supplementary Section 8, the *z*-components of SAM contain both L-spins and helicity-dependent T-spins. These T-spins originate from the helicity-related kinetic momentum densities in the *xy*-plane and this kinetic momentum would be inverted when the incident optical helicity is reversed. Moreover, their spatial distributions would be inverted to those of L-spins as evident from the theoretical analysis given in the previous paragraph. By further considering the reversal of the propagating direction, four momentum-locked T-spin states are found in the focused CP light systems, which is consistent with the nontrivial spin Chern number of an optical wave-packet and reveals these T-spins to possess $\mathbb{Z}4$ topological invariance.

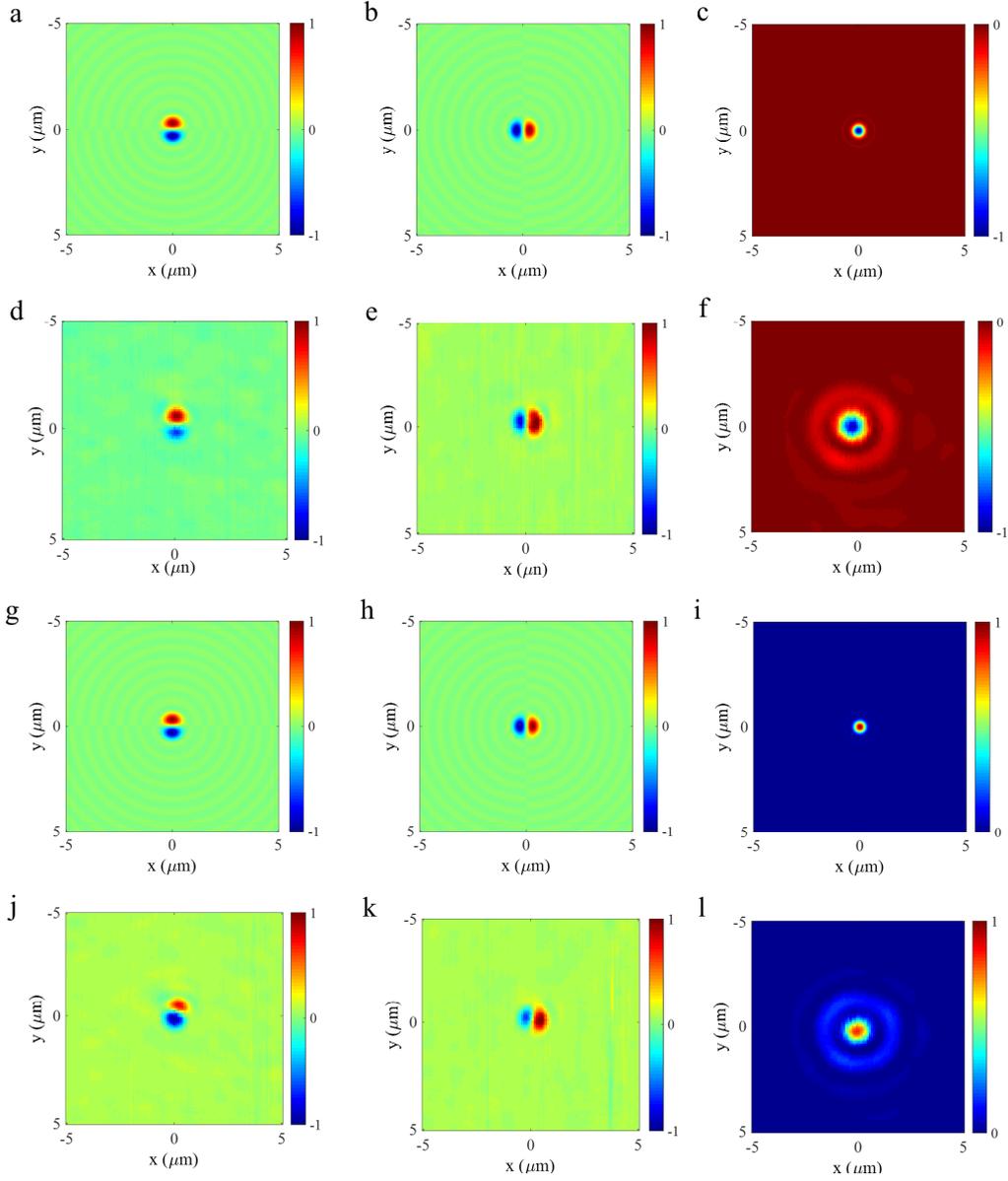

**Fig. 4. Experimental validation of the T-spin and spin-momentum locking in a focused circularly polarized beam (CPB):** (a-c) Theoretically calculated *x*-, *y*-, and *z*-components, respectively, of the SAM densities of a focused LCP beam. (d-f), Corresponding experimental results. (g-i), Same as (a-c), but with a RCP beam. (j-l), Corresponding experimental results for the RCP beam. The optical axis is along the z-direction. As the incident beam is changed from LCP to RCP, the *x*- and *y*-components of SAM remain unchanged, thereby manifesting a helicity-independent T-spin. In contrast, the sign of the z-component of the SAM changes from positive to negative. This spin component contains two parts: the L-spin and the helicity-dependent T-spin. Together with the inverted propagating property, the T-spin of the focused circular polarized light was demonstrated to possess $\mathbb{Z}4$ topological invariance, which matches well with the theoretical analysis.

**Discussions and conclusions:** To summarize, from theory, we derived a unified theory that involves the decomposition of EM spin and uncovered the underlying physical difference between T-spin and L-spin. The L-spin is solely determined by the EM helicity but needs to consider coupling effects, whereas T-spin originates from the spatial inhomogeneity of the kinetic momentum density and undergoes a universal spin-momentum locking. Here, we emphasize that the T-spin is locked with the kinetic momentum rather than with the canonical momentum given by the mean wavevector. Indeed, the T-spin can be oriented parallel to the mean wavevector. The T-spin further decomposes into helicity-independent and helicity-dependent components, which are determined separately by the transverse gradient of a helicity-unrelated and a helicity-related kinetic momentum. Thus, there exist four spin-momentum locking states, the number being consistent with the nontrivial topological spin Chern number. Moreover, the T-spin bearing the curl-relationship with the kinetic momentum is closely related to the Berry curvature of an EM system. Specially, the helicity-dependent T-spin which is associated with the inverted helical component can be explained based on the evolution of the geometric phase in EM systems.

Our theory has an interdisciplinary impact and is extendible to other classical wave fields. For example, the spin-momentum locking relationship of the longitudinal acoustic wave (identified with subscript A) can be expressed as $\mathbf{S}_A = \nabla \times \mathbf{\Pi}_A / k_A^2$, and for a deep-water gravity water wave (identified with subscript G) [6], a similar spin-momentum relationship $\mathbf{S}_G = \nabla_2 \times \mathbf{\Pi}_G / 2k_G^2$ obtains. We noted that the longitudinal spin is absent because longitudinal acoustic waves and surface water waves can be considered as spin-0 phonons. These results reveal the physical origins and topological properties of spin in diverse classical wave fields and illuminate the universality of spin-momentum locking. They motivate explorations of field theory based on spin degrees of freedom and constructions of novel chiral spin textures [46,47]. For a clear comparison, we have summarized in Table 1 the primary dynamical properties of these three types of wave fields.

For applications, this spin-momentum locking property of T-spin in a generic EM field can be utilized to construct the diverse photonic topological spin structures, such as the Néel-type skyrmions in optical near fields and Bloch-type skyrmions in free space [20]. Moreover, we performed numerical simulations (Supplementary Section 11) to demonstrate properties of momentum-locked transverse optical forces by considering the interaction between metallic helical nanostructures and Bloch-type skyrmions in free space. These simulations suggest further applications in chiral sorting using photonic topological spin structures. Overall, the findings reveal a unified field theory to describe the spin–orbit coupling of light based on the spin degrees of freedom and wave–matter interactions in interdisciplinary research and motivates explorations of novel applications in optical manipulation, chiral quantum optics, and electronics [26,48,49].

**Materials and Methods**

**Experimental setup.** The experimental setup for studies of the optical spin-momentum locking is shown in Supplementary Section 9, **Fig. S10**. The experiment was performed on the example of surface plasmon polaritons (SPPs), which are TM mode evanescent waves supported at a metal-dielectric interface. A He-Ne laser beam with a wavelength of 632.8nm was used as a light source. After a telescope system to expand the beam, a combination of linear polarizer (LP), half-wave plates (HWP), quarter-wave plates (QWPs) and vortex wave plates (VWPs) was employed to modulate the polarization of the laser beam. A spatial light modulator (SLM) was then utilized to modulate the phase of the beam. The structured beam was then tightly focused by an oil-immersion objective (Olympus, NA=1.49, 100×) onto the sample consisting of a thin silver film (45-nm thickness) deposited on a cover slip, to form the desired SPP beams at the air/silver interface.

The experimental setup for mapping the SAM components perpendicular to optical axis (Supplementary Section 9, **Fig. S10**) comprises an incident beam (wavelength: 632.8 nm) that is tightly focused by an objective lens (Olympus, NA=0.5, 50×) onto a PS nanoparticle (diameter: 201 nm) sitting on a silver film (thickness: 45 nm). The focusing field and the scattering field of the PS particle (the far-field radiation field and part of the near-field evanescent field) radiate downward by coupling with the silver film. The signal is collected by an oil-immersed objective lens (Olympus, NA = 1.49, 100×). Using a high-precision piezo-stage (Physik Instrumente, P-545), we moved the PS particle through the focal plane of the tightly focused beam. Each time the position is moved, the back focal plane intensity (far-field intensity) distribution is imaged using a four-quadrant detector. From dipole theory and similar techniques described in Ref. (10), the transverse components of the SAM density can be reconstructed.

The setup of the tip-fibre-based-measurement system to map the SAM component parallel to optical axis [Supplementary Section 10, **Fig. S13(a)**] comprises a He–Ne laser (operating wavelength: 632.8 nm) used as a light source. The light beam is expanded and collimated via a telescope system and then passed through a linear polarizer (LP) and a quarter wave plate (QWP) to produce the desired LCP or RCP light. The beam is then focused using an objective lens (Olympus, NA 0.7, 60×) onto a silica coverslip for further image scanning by a self-assembly near-field scanning optical microscopic system. The system's probe has a nanohole and is controlled using a tuning fork feedback system for mapping the in-plane field distributions of the focused beams. The near-field signal, which couples via the nanohole to the fibre, is split and then analysed using a combination of QWP and LP to extract the individual circular polarization components of the signal ($I_{\text{LCP}}$: LCP component and $I_{\text{RCP}}$: RCP component). These components are then directed to two photomultiplier tubes (PMTs) to measure the intensity of the two signals. This then enables a quantification of the out-of-plane SAM component (i.e., along the optical axis) of the focused beams using the relation,

$$S_z = \frac{\varepsilon}{4\omega} \frac{k^2 + \kappa^2}{\kappa^2} \left( I_{\text{RCP}} - I_{\text{LCP}} \right).$$

### Acknowledgements

This work was supported, in part, by the Guangdong Major Project of Basic Research grant 2020B0301030009, the National Natural Science Foundation of China grants 61935013, 62075139, 61427819, 61622504, 12174266, the Leadership of Guangdong province program grant 00201505, and the Science and Technology Innovation Commission of Shenzhen grants JCYJ20200109114018750. L.D. acknowledges the support given by the Guangdong Special Support Program.

### Author contributions

All authors contributed to the article.

### Competing interests

The authors declare no competing interests.

### Data availability

The data that support the plots within this paper and other findings of this study are available from the corresponding author upon reasonable request.

Supplementary Materials for

"Spin decomposition and topological properties of optical spin angular momenta in general electromagnetic fields"

Peng Shi, Luping Du*, Aiping Yang, Xiaojin Yin, Xinrui Lei, Xiaocong Yuan*

Nanophotonics Research Centre, Shenzhen Key Laboratory of Micro-Scale Optical Information Technology, Institute of micro/nano optoelectronics, Shenzhen University, Shenzhen, 518060, China

*Authors to whom correspondence should be addressed: *lpdu@szu.edu.cn* and *xcyuan@szu.edu.cn.*

**Contents:**

**Supplementary Section 1:** Basic dynamical physical quantities in spin-orbit interaction of light

To investigate the spin-orbit interaction of a time-harmonic monochromatic electromagnetic (EM) field (**E**: electric field; **H**: magnetic field) with angular frequency $\omega$ in a lossless isotropic medium with permittivity $\varepsilon$ and permeability $\mu$, we first introduce the expressions of the cycle-averaged energy density $W$, kinetic-Abraham-Poynting momentum density **Π**, canonical momentum density **P** and total spin angular momentum (SAM) density **S** as follows (SI units are used through the manuscript and supplementary materials) (S1):

$$W = \langle \psi | \psi \rangle = \frac{1}{4}\left\{ \varepsilon \mathbf{E}^* \cdot \mathbf{E} + \mu \mathbf{H}^* \cdot \mathbf{H} \right\}, \tag{S1}$$

$$\mathbf{P} = \frac{1}{\hbar\omega}\langle \psi | \hat{\mathbf{P}} | \psi \rangle = \frac{1}{4\omega}\operatorname{Im}\left\{ \varepsilon \mathbf{E}^* \cdot (\nabla) \mathbf{E} + \mu \mathbf{H}^* \cdot (\nabla) \mathbf{H} \right\}, \tag{S2}$$

$$\mathbf{\Pi} = \langle \psi | \hat{\boldsymbol{\tau}}/v | \psi \rangle = \frac{1}{2v^2}\operatorname{Re}\left\{ \mathbf{E}^* \times \mathbf{H} \right\}, \tag{S3}$$

and

$$\mathbf{S} = \frac{1}{\hbar\omega}\langle \psi | \hat{\mathbf{S}} | \psi \rangle = \frac{1}{4\omega}\operatorname{Im}\left\{ \varepsilon \left(\mathbf{E}^* \times \mathbf{E}\right) + \mu \left(\mathbf{H}^* \times \mathbf{H}\right) \right\}. \tag{S4}$$

Here, the 6-vector $|\psi\rangle = \left[\sqrt{\varepsilon}\mathbf{E}, i\sqrt{\mu}\mathbf{H}\right]^T/2$ represents the photon wave function (S2, S3) and the symbol T represents the transpose of matrix; $\mathbf{X} \cdot (\nabla)\mathbf{Y} \equiv \Sigma_i x_i \nabla y_i$; $\hat{\mathbf{P}} = -i\hbar\nabla$ denotes the momentum operator in quantum mechanics; the symbol $\hat{\boldsymbol{\tau}}/v = \left[0, \hat{\mathbf{S}}/v\, ; \hat{\mathbf{S}}/v\, , 0\right]$ denotes the kinetic momentum operator with $v = 1/\sqrt{\varepsilon\mu}$ being the velocity of light in medium and $\hat{\mathbf{S}}$ is the spin-1 matrix in SO(3) (46); $\hbar$ is the reduced Plank constant and the symbol $*$ indicates the complex conjugate. Here, we ignore the dispersion through the manuscript and supplemental materials, which will induce an additional group term in the permittivity and permeability (S1).

From the spin-orbit decomposition proposed by M. V. Berry (46), the kinetic momentum **Π** can be decomposed into the orbital momentum **P** (which is also regarded as the canonical momentum) and Belinfante spin momentum ($\mathbf{P}_s = \nabla \times \mathbf{S}/2$): **Π** = **P** + **P**$_s$. Therein, the orbital/canonical momentum **P** can naturally be associated with the local wavevector of a single EM plane wave or the mean wavevector (propagating direction) of the total structured EM field as $\bar{\mathbf{k}} = \mathbf{P}/\hbar$. (Thus, for the sake of rigorous, we call the $\bar{\mathbf{k}}$ of plane wave as local wavevector and the $\bar{\mathbf{k}}$ of structured light as mean wavevector.) The directional vector is $\hat{\mathbf{k}} = \bar{\mathbf{k}}/k$ and the modulus is $k = |\bar{\mathbf{k}}| = \omega/\sqrt{\varepsilon\mu}$ (51).

Naturally, in the passive homogeneous medium, there are:

$$\nabla \cdot \mathbf{\Pi} = 0, \quad \nabla \cdot \mathbf{P} = 0, \quad \nabla \cdot \mathbf{P}_s = 0, \tag{S5}$$

$$\nabla \cdot \mathbf{S} = 0, \tag{S6}$$

$$\mathbf{\Pi} = \mathbf{P} + \mathbf{P}_s, \tag{S7}$$

and

$$\mathbf{P}_s = \nabla \times \mathbf{S}/2. \tag{S8}$$

The former equations (S5-S8) have been proposed in plentiful references (10, 28, 56). Here, we aim to investigate the spin decomposition into the longitudinal spin (L-spin) and the transverse spin (T-spin) for a generic EM field (akin to the Helmholtz decomposition of an arbitrary vector mathematically) in free space or at the interface, and the resulted spin-momentum topological properties.

**Supplementary Section 2:** Spin properties of elliptically polarized propagating plane waves

In this section, we investigate the dynamical properties of propagating EM plane waves. The dynamical properties of the propagating EM plane waves have been studied by many researchers (28, 30) and thus we directly summarize the results here.

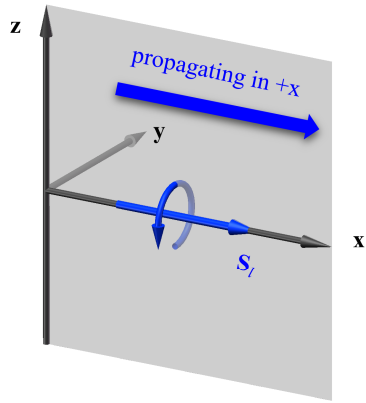

**Fig. S1.** The schematic diagram of elliptically polarized plane wave propagating in +$x$-direction. Only the L-spin exists in the case.

Assuming the electric field components of a plane wave propagating in the $x$-direction (in **Fig. S1**) at an orthogonal coordinate system ($x$, $y$, $z$) as:

$$\mathbf{E}(\mathbf{r}) = \left(0\hat{\mathbf{x}} + A_s\hat{\mathbf{y}} + A_p\hat{\mathbf{z}}\right)e^{i(kx-\omega t)},$$

the magnetic field components are:

$$\mathbf{H}(\mathbf{r}) = \left(0\hat{\mathbf{x}} - A_p\hat{\mathbf{y}} + A_s\hat{\mathbf{z}}\right)e^{i(kx-\omega t)}/\eta,$$

where $\eta = \sqrt{\mu/\varepsilon}$ is the wave impedance, $A_s$, $A_p$ are the complex amplitudes of $s$-polarization and $p$-polarization components, respectively. Since the EM wave considered here is time-harmonic throughout, we ignore the $\exp(-i\omega t)$ in the following expressions. From the expressions of electric field and magnetic field, it can be deduced that the energy density and the canonical momentum are

$$W = \frac{1}{4}\left\{\varepsilon|\mathbf{E}|^2 + \mu|\mathbf{H}|^2\right\} = \frac{\varepsilon}{2}\left\{A_s^*A_s + A_p^*A_p\right\}, \tag{S9}$$

$$\mathbf{P} = \frac{1}{4\omega}\text{Im}\left\{\varepsilon\mathbf{E}^*\cdot(\nabla)\mathbf{E} + \mu\mathbf{H}\cdot(\nabla)\mathbf{H}\right\} = \frac{W}{\hbar\omega}\left(\hbar k\hat{\mathbf{x}} + 0\hat{\mathbf{y}} + 0\hat{\mathbf{z}}\right), \tag{S10}$$

respectively. The local wavevector associated with the canonical momentum is

$$\overline{\mathbf{k}} = \frac{\mathbf{P}}{\hbar} = \frac{W}{\hbar\omega}(k\hat{\mathbf{x}} + 0\hat{\mathbf{y}} + 0\hat{\mathbf{z}}),\tag{S11}$$

and its unit directional vector is

$$\hat{\mathbf{k}} = \frac{\overline{\mathbf{k}}}{k} = \frac{W}{\hbar\omega}(1\hat{\mathbf{x}} + 0\hat{\mathbf{y}} + 0\hat{\mathbf{z}}).\tag{S12}$$

In this way, the canonical momentum can be rewritten as $\mathbf{P} = \hbar\overline{\mathbf{k}} = \hbar k\hat{\mathbf{k}}$. Then, the SAM is

$$\mathbf{S} = \frac{W}{\omega}\left(\frac{\mathrm{Im}\{A_s^* A_p - A_p^* A_s\}}{A_s^* A_s + A_p^* A_p}\hat{\mathbf{x}} + 0\hat{\mathbf{y}} + 0\hat{\mathbf{z}}\right) = \hbar\sigma\frac{\overline{\mathbf{k}}}{k} = \hbar\sigma\hat{\mathbf{k}}.\tag{S13}$$

Equation (S13) shows that the spin vector is parallel to the local wavevector (as shown in **Fig. S1**), and thus previously, the researchers named this type of spin as L-spin ($\mathbf{S}_l$). Here, the EM helicity $\sigma$ is defined as (14, 28, 30)

$$\sigma = \frac{\mathrm{Im}\{A_s^* A_p - A_p^* A_s\}}{A_s^* A_s + A_p^* A_p}.\tag{S14}$$

which also represents the polarization ellipticity of EM field.

The kinetic momentum is calculated as

$$\mathbf{\Pi} = \frac{\varepsilon\mu}{2}\mathrm{Re}\{\mathbf{E}^* \times \mathbf{H}\} = \frac{W}{\omega}(k\hat{\mathbf{x}} + 0\hat{\mathbf{y}} + 0\hat{\mathbf{z}}).\tag{S15}$$

It can be observed that the time-averaging kinetic momentum of this propagating plane wave is homogeneous in space, and the T-spin given by the vorticity of kinetic momentum is

$$\mathbf{S}_t = \frac{1}{2k^2}\nabla \times \mathbf{\Pi} = 0.\tag{S16}$$

In summary, the spin-momentum properties of elliptically polarized plane waves can be summed as:

$$\mathbf{S}_l = \hbar\sigma\hat{\mathbf{k}},\ \mathbf{S}_t = \frac{1}{2k^2}\nabla \times \mathbf{\Pi} = 0\ \text{and}\ \mathbf{S} = \mathbf{S}_l + \mathbf{S}_t,\tag{S17}$$

which coincides well with the former results in several review papers (10, 28, 30, 56).

**Supplementary Section 3:** Spin properties of elliptically polarized evanescent plane waves

In the section, we investigate the spin-momentum properties of evanescent plane waves. We assume the interface is at $z = 0$ and we only consider the evanescent modes at $z > 0$ since the evanescent modes at $z < 0$ has the similar spin-momentum properties (18, 19). The electric field components of an elliptically polarized evanescent plane wave (decaying exponentially in the $+z$-direction) can be given by:

$$\mathbf{E} = \left(-A_p\frac{i\kappa}{k}\hat{\mathbf{x}} + A_s\hat{\mathbf{y}} + A_p\frac{k_p}{k}\hat{\mathbf{z}}\right)\exp(ik_p x - \kappa z)\tag{S18}$$

as shown in **Fig. S2(A)**. Obviously, the electric field satisfies the Gauss' law:

$$\nabla \cdot \mathbf{E} = \left(-A_p\frac{i\kappa}{k}ik_p - A_p\frac{k_p}{k}\kappa\right)\exp(ik_p x - \kappa z) = 0.\tag{S19}$$

Here, $A_s$, $A_p$ are the complex amplitudes of s-polarization and p-polarization components, respectively; $k$, $k_p$ and $i\kappa$ are the total, horizontal and normal wavevectors, respectively; the wavevectors satisfy the relation: $k_p^2 = k^2 + \kappa^2$. From the expression of electric field, one can obtain the magnetic field as:

$$\mathbf{H} = \frac{1}{\eta}\left(-A_s \frac{i\kappa}{k}\hat{\mathbf{x}} - A_p \hat{\mathbf{y}} + A_s \frac{k_p}{k}\hat{\mathbf{z}}\right)\exp(ik_p x - \kappa z). \quad (S20)$$

From the electric and magnetic field components, the energy density and the canonical momentum are:

$$W = \frac{\varepsilon}{2}\frac{k_p^2}{k^2}\{A_s^* A_s + A_p^* A_p\}e^{-2\kappa z}, \quad (S21)$$

$$\mathbf{P} = \frac{\varepsilon}{2\omega}\frac{k_p^3}{k^2}\left(\{A_s^* A_s + A_p^* A_p\}\hat{\mathbf{x}} + 0\hat{\mathbf{y}} + 0\hat{\mathbf{z}}\right)e^{-2\kappa z} = \frac{W}{\omega}\left(k_p\hat{\mathbf{x}} + 0\hat{\mathbf{y}} + 0\hat{\mathbf{z}}\right), \quad (S22)$$

respectively. Both the energy density and the canonical momentum decay exponentially in the +z-direction. Then, the local wavevector of the evanescent plane wave is

$$\bar{\mathbf{k}} = \frac{\mathbf{P}}{\hbar} = \frac{W}{\hbar\omega}\left(k_p\hat{\mathbf{x}} + 0\hat{\mathbf{y}} + 0\hat{\mathbf{z}}\right), \quad (S23)$$

and its unit directional vector is

$$\hat{\mathbf{k}} = \frac{\bar{\mathbf{k}}}{k} = \frac{W}{\hbar\omega}\frac{k_p}{k}\left(1\hat{\mathbf{x}} + 0\hat{\mathbf{y}} + 0\hat{\mathbf{z}}\right). \quad (S24)$$

Similarly, the canonical momentum of the evanescent plane wave can be rewritten as $\mathbf{P} = \hbar\bar{\mathbf{k}} = \hbar k\hat{\mathbf{k}}$.

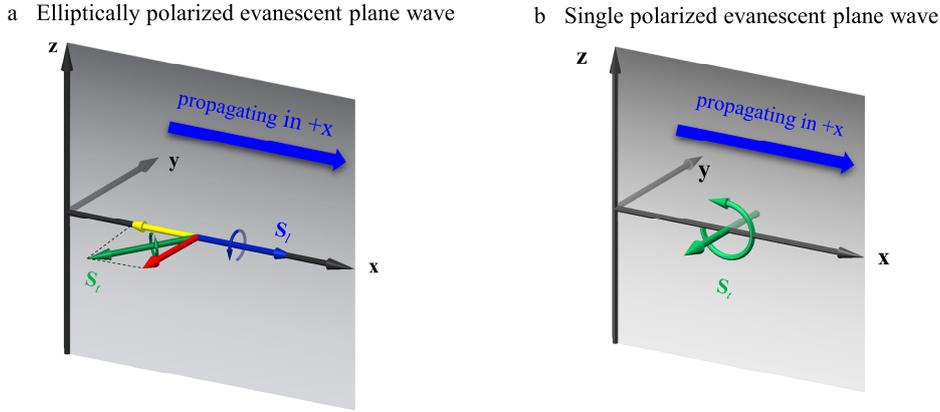

**Fig. S2.** (a) The schematic diagram of spin property of an elliptically polarized evanescent plane wave propagating in x-direction at an interface $z = 0$. The L-spin and T-spin exist simultaneously in the case. (b) The schematic diagram of spin property of a single polarized evanescent plane wave propagating in x-direction at an interface $z = 0$. Only the helicity-independent T-spin exists in the case.

Subsequently, the SAM can be calculated to be

$$\mathbf{S} = \frac{W}{\omega}\frac{\kappa}{k_p}\left(\frac{k}{\kappa}\frac{\text{Im}\{A_s^* A_p - A_s A_p^*\}}{A_s^* A_s + A_p^* A_p}\hat{\mathbf{x}} - 1\hat{\mathbf{y}} + 0\hat{\mathbf{z}}\right) = \frac{k^2}{k_p^2}\hbar\sigma\hat{\mathbf{k}} - \frac{W}{\omega}\frac{\kappa}{k_p}\hat{\mathbf{y}} + 0\hat{\mathbf{z}}. \quad (S25)$$

with its EM helicity

$$\sigma = \frac{\text{Im}\{A_s^* A_p - A_s A_p^*\}}{A_s^* A_s + A_p^* A_p}.$$

Previously, most researches (10, 15, 28, 30) regarded the *y*-component SAM, which is helicity-independent, as the T-spin. This is correct for the single polarized mode ($A_p = 0$ or $A_s = 0$), as shown in **Fig. S2(b)**. However, it is incomplete for the elliptically polarized evanescent plane wave as shown in **Fig. S2(a)**.

In the following, we calculate the kinetic momentum and the T-spin given by the vorticity of the kinetic momentum as:

$$\mathbf{\Pi} = \frac{W}{c}\frac{k}{k_p}\left(1\hat{\mathbf{x}} - \frac{\kappa}{k}\sigma\hat{\mathbf{y}} + 0\hat{\mathbf{z}}\right) = \hbar\frac{k^3}{k_p^2}\left(1\hat{\mathbf{k}} - \frac{W}{\hbar\omega}\frac{\kappa k_p}{k^2}\sigma\hat{\mathbf{y}} + 0\hat{\mathbf{z}}\right), \quad (S26)$$

and

$$\mathbf{S}_t = \frac{1}{2k^2}\nabla\times\mathbf{\Pi} = \frac{W}{\omega}\frac{\kappa}{k_p}\left(-\frac{\kappa}{k}\sigma\hat{\mathbf{p}} - 1\hat{\mathbf{s}} + 0\hat{\mathbf{z}}\right). \quad (S27)$$

Equation (S26) indicates that the kinetic momentum has two components: the component that is along the local wavevector is helicity-irrelevant and the *y*-component is helicity-related. This helicity-related kinetic momentum can be found widely in paraxial system (the elliptically polarized Gaussian light) and nonparaxial system (the tightly focusing of the elliptically polarized plane wave). In the case, the spin momentum does not parallel to canonical momentum because the spin momentum has *y*-component. In addition, Eq. (S27) shows that the T-spin has two components: the *y*-component is helicity-independent and the *x*-component is helicity-dependent.

Then, by subtracting the vorticity of kinetic momentum (S27) from the total SAM **S** (S25), we obtain:

$$\mathbf{S}_l = \mathbf{S} - \frac{1}{2k^2}\nabla\times\mathbf{\Pi} = \frac{W}{\omega}\frac{k_p}{k}\left(\frac{\text{Im}\{A_s^* A_p - A_s A_p^*\}}{|A_p|^2 + |A_s|^2}\hat{\mathbf{p}} + 0\hat{\mathbf{s}} + 0\hat{\mathbf{z}}\right) = \hbar\sigma\hat{\mathbf{k}}. \quad (S28)$$

This term is consistent with the L-spin of an elliptically polarized propagating plane wave, which indicates this remaining SAM is the L-spin. This consistency can be explained as: in the interface system, there is conservation of *z*-component total angular momentum instead of total SAM owing to the breaking of inverse symmetry. However, since all the canonical momentum, the local wavevector and the SAM are normalized by the energy density, the aforementioned spin-momentum properties can also be referred to a single wave-packet. By properly choosing the coordinates to eliminate the extrinsic orbital angular momentum (10), there will be conservation of intrinsic SAM of a single wave-packet when the intrinsic orbital angular momentum (vortex phase) is absent, although the total SAM is never conservative in the nonparaxial system due to the spin-orbit coupling (10).

In summary, the spin-momentum properties of elliptically polarized evanescent plane waves can be summed as:

$$\mathbf{S}_l = \hbar\sigma\hat{\mathbf{k}}, \quad \mathbf{S}_t = \frac{1}{2k^2}\nabla\times\mathbf{\Pi} = \frac{W}{\omega}\frac{\kappa}{k_p}\left(-\frac{\kappa}{k}\sigma\hat{\mathbf{p}} - 1\hat{\mathbf{s}} + 0\hat{\mathbf{z}}\right) \text{ and } \mathbf{S} = \mathbf{S}_l + \mathbf{S}_t. \quad (S29)$$

In addition, equation (S29) represents the spin-momentum locking property in classical EM system. The effect of intrinsic spin-momentum locking originated from the intrinsic spin-orbit coupling of Maxwell's equations has been investigated intensively by many research groups (30-36). However, all the former works put emphasis on the spin-momentum locking in single polarized modes. Here, from equation (S29), it can be found that the spin-momentum locking is also helicity-dependent, as shown in **Fig. S3**. There are two pairs of spin-momentum locking states, which are consistent with the nontrivial photonic spin Chern number and indicates the light satisfy the $\mathbb{Z}_4$ topological invariance (16, 20).

In the special case that the dual symmetry between the electric and magnetic properties is broken, there is only one polarized state survive at the optical interface (such as: surface plasmons polaritons (S4) or Bloch surface wave (S5) as shown in **Fig. S2(b)**). The results can be downgraded into:

$$\mathbf{S}_l = 0, \quad \mathbf{S}_t = \frac{1}{2k^2}\nabla \times \mathbf{\Pi} = \frac{W}{\omega}\frac{\kappa}{k_p}(0\hat{\mathbf{x}} - 1\hat{\mathbf{y}} + 0\hat{\mathbf{z}}) \quad \text{and} \quad \mathbf{S} = \mathbf{S}_l + \mathbf{S}_t, \tag{S30}$$

which indicate the intrinsic spin-momentum locking of single polarized modes (16-20).

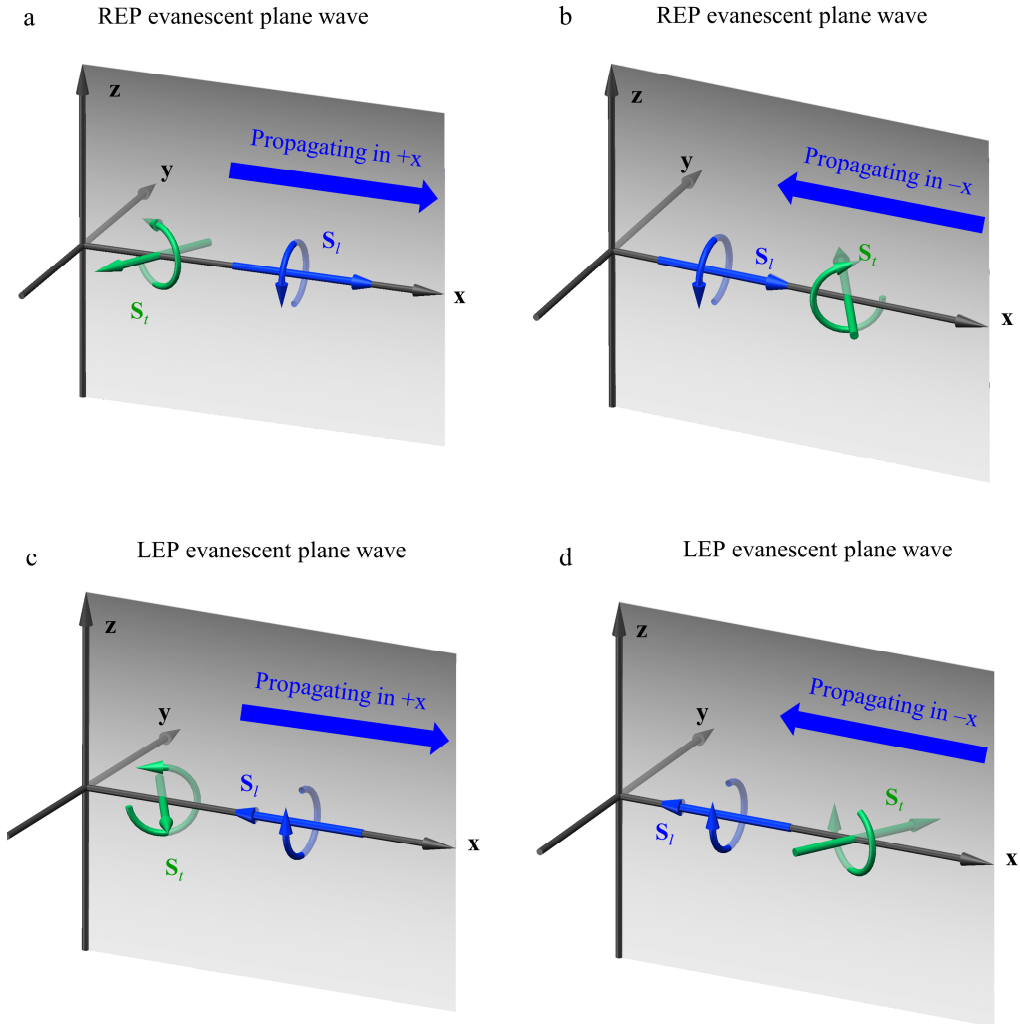

**Fig. S3.** Spin-momentum locking states of the helicity-dependent T-spins of elliptically polarized evanescent plane waves. The L-spin and T-spin of right-handed elliptically polarized (REP) evanescent plane wave propagating in the (a) +x-direction and (b) –x-direction; the L-spin and T-spin of left-handed

elliptically polarized (LEP) evanescent plane wave propagating in the (c) +x-direction and (d) –x-direction. The x component T-spin originated from the inhomogeneity of helicity-related kinetic momentum perpendicular to the local wavevector is helicity-dependent and inverted to the direction of L-spin, whereas the s component T-spin originated from the inhomogeneity of helicity-irrelevant kinetic momentum along the local wavevector is helicity-independent.

**Supplementary Section 4:** Spin properties of interfering evanescent plane waves

In this section, we investigate the spin properties of the generic evanescent field. From the superposition theory of states (45), an arbitrary structured light can be expanded into the superposition of plane wave basis. Here, for the sake of simplicity, we only consider the two-waves interference, but the results can be extended to the situation of the multi-waves interference.

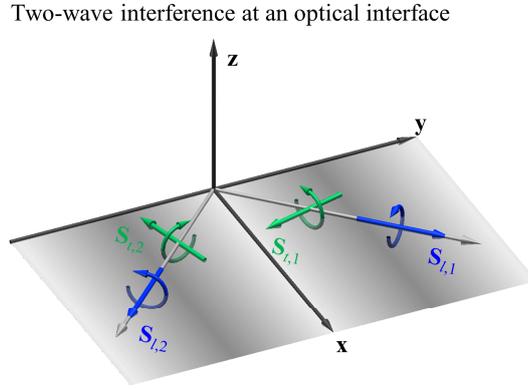

**Fig. S4.** The schematic diagram of the interference between two elliptically polarized evanescent plane waves. The intersected angle between the individual wave and the *x*-axis is $\Theta_1$.

The two elliptically polarized evanescent plane waves (waves 1: $A_{s1}$, $A_{p1}$; wave 2: $A_{s2}$, $A_{p2}$) propagate in the *xy*-plane with an intersected angle $2\Theta_1$ between their wave vectors

$$\mathbf{k}_{1,2} = k_{p1}\hat{\mathbf{x}} \pm k_{s1}\hat{\mathbf{y}} - \kappa\hat{\mathbf{z}},$$

where $k_p^2 = k^2 + \kappa^2$, $k_{p1} = k_p\cos\Theta_1$ and $k_{s1} = k_p\sin\Theta_1$ as shown in **Fig. S4**. The interface is at $z = 0$, and only the evanescent modes in the region $z > 0$ is considered here.

For the individual evanescent plane wave, the energy density, directional vector of local wave vector, and L-spin can be expressed as:

$$W_1 = \frac{\varepsilon}{2}\frac{k_p^2}{k^2}\{A_{p1}^*A_{p1} + A_{s1}^*A_{s1}\}e^{-2\kappa z} \quad W_2 = \frac{\varepsilon}{2}\frac{k_p^2}{k^2}\{A_{p2}^*A_{p2} + A_{s2}^*A_{s2}\}e^{-2\kappa z}, \tag{S31}$$

$$\hat{\mathbf{k}}_1 = \frac{W_1}{\hbar\omega}\frac{k_p}{k}\left(k_{p1}/k_p\,\hat{\mathbf{x}} + k_{s1}/k_p\,\hat{\mathbf{y}} + 0\hat{\mathbf{z}}\right) \quad \hat{\mathbf{k}}_2 = \frac{W_2}{\hbar\omega}\frac{k_p}{k}\left(k_{p2}/k_p\,\hat{\mathbf{x}} - k_{s2}/k_p\,\hat{\mathbf{y}} + 0\hat{\mathbf{z}}\right), \tag{S32}$$

$$\mathbf{S}_{l,1} = \hbar\frac{\text{Im}\{A_{s1}^*A_{p1} - A_{p1}^*A_{s1}\}}{A_{p1}^*A_{p1} + A_{s1}^*A_{s1}}\hat{\mathbf{k}}_1 = \hbar\sigma_1\hat{\mathbf{k}}_1 \quad \mathbf{S}_{l,2} = \hbar\frac{\text{Im}\{A_{s2}^*A_{p2} - A_{p2}^*A_{s2}\}}{A_{p2}^*A_{p2} + A_{s2}^*A_{s2}}\hat{\mathbf{k}}_2 = \hbar\sigma_2\hat{\mathbf{k}}_2. \tag{S33}$$

These results match well with those of Equations (S21), (S24) and (S28).

Subsequently, for the interference field, the energy density is

$$W = \frac{\varepsilon}{2}\frac{k_p^2}{k^2}\left\{ + \underbrace{A_{s1}^*A_{s1} + A_{p1}^*A_{p1}}_{W_1} + \underbrace{A_{s2}^*A_{s2} + A_{p2}^*A_{p2}}_{W_2} + \underbrace{\frac{k_{p1}^2}{k_p^2}\begin{pmatrix} +A_{s1}^*A_{s2}e^{-2ik_{s1}y} + A_{s1}A_{s2}^*e^{2ik_{s1}y} \\ +A_{p1}^*A_{p2}e^{-2ik_{s1}y} + A_{p1}A_{p2}^*e^{2ik_{s1}y} \end{pmatrix}}_{W_c} \right\} e^{-2\kappa z}$$ (S34)

$$= W_1 + W_2 + W_c$$

Here, the term $W_c$ is originated from the coupling between the two evanescent plane waves and is considered as the crossed energy density.

The total SAM and the T-spin given by the vorticity of kinetic momentum are

$$\mathbf{S} = \frac{\varepsilon e^{-2\kappa z}}{4\omega}\operatorname{Im}\left\{ \begin{pmatrix} \left\{ +\frac{i\kappa}{k}\frac{k_{s1}}{k}\left[A_{s1}^*A_{s1} + A_{p1}^*A_{p1} - A_{s2}^*A_{s2} - A_{p2}^*A_{p2}\right] \\ +\frac{k_{p1}}{k}\begin{bmatrix} +A_{s1}^*A_{p1} - A_{s1}A_{p1}^* + A_{s2}^*A_{p2} - A_{s2}A_{p2}^* \\ +A_{s1}^*A_{p2}e^{-2ik_{s1}y} - A_{s1}A_{p2}^*e^{+2ik_{s1}y} + A_{s2}^*A_{p1}e^{+2ik_{s1}y} - A_{s2}A_{p1}^*e^{-2ik_{s1}y} \end{bmatrix} \right\}\hat{\mathbf{x}} \\ \left\{ -\frac{i\kappa}{k}\frac{k_{p1}}{k}\begin{bmatrix} +A_{s1}^*A_{s1} + A_{p1}^*A_{p1} + A_{s2}^*A_{s2} + A_{p2}^*A_{p2} \\ +A_{s2}^*A_{s1}e^{+2ik_{s1}y} + A_{s1}^*A_{s2}e^{-2ik_{s1}y} \\ +A_{p2}^*A_{p1}e^{+2ik_{s1}y} + A_{p1}^*A_{p2}e^{-2ik_{s1}y} \end{bmatrix} + \frac{k_{s1}}{k}\begin{bmatrix} A_{s1}^*A_{p1} - A_{s1}A_{p1}^* \\ -A_{s2}^*A_{p2} + A_{s2}A_{p2}^* \end{bmatrix} \right\}\hat{\mathbf{y}} \\ \left\{ +\frac{k_{p1}}{k}\frac{k_{s1}}{k}\begin{bmatrix} +A_{s2}^*A_{s1}e^{+2ik_{s1}y} - A_{s1}^*A_{s2}e^{-2ik_{s1}y} \\ +A_{p2}^*A_{p1}e^{+2ik_{s1}y} - A_{p1}^*A_{p2}e^{-2ik_{s1}y} \end{bmatrix} \right\}\hat{\mathbf{z}} \end{pmatrix} \right\}, \quad (S35)$$

and

$$\mathbf{S}_t = \frac{1}{2k^2}\nabla\times\mathbf{\Pi} = \frac{\varepsilon e^{-2\kappa z}}{2\omega}\begin{pmatrix} \left\{ -\frac{\kappa^2}{k}\frac{k_{p1}}{k}\operatorname{Im}\begin{bmatrix} +A_{s1}^*A_{p1} - A_{s1}A_{p1}^* \\ +A_{s2}^*A_{p2} - A_{s2}A_{p2}^* \end{bmatrix} + \frac{\kappa}{k}\frac{k_{s1}}{k}\begin{bmatrix} +A_{s1}^*A_{s1} + A_{p1}^*A_{p1} \\ -A_{s2}^*A_{s2} - A_{p2}^*A_{p2} \end{bmatrix} \\ +\frac{k_{s1}^2-\kappa^2}{k^2}\frac{k_{p1}}{k}\operatorname{Im}\begin{bmatrix} +A_{s1}^*A_{p2}e^{-2ik_{s1}y} - A_{s1}A_{p2}^*e^{+2ik_{s1}y} \\ +A_{s2}^*A_{p1}e^{+2ik_{s1}y} - A_{s2}A_{p1}^*e^{-2ik_{s1}y} \end{bmatrix} \right\}\hat{\mathbf{x}} \\ \left\{ -\frac{\kappa^2}{k^2}\frac{k_{s1}}{k}\operatorname{Im}\begin{bmatrix} +A_{s1}^*A_{p1} - A_{s1}A_{p1}^* \\ -A_{s2}^*A_{p2} + A_{s2}A_{p2}^* \end{bmatrix} - \frac{\kappa}{k}\frac{k_{p1}}{k}\begin{bmatrix} +A_{s1}^*A_{s1} + A_{p1}^*A_{p1} \\ +A_{s2}^*A_{s2} + A_{p2}^*A_{p2} \end{bmatrix} \\ -\frac{\kappa}{k}\frac{k_{p1}}{k}\begin{bmatrix} +A_{s1}^*A_{s2}e^{-2ik_{s1}y} + A_{s2}^*A_{s1}e^{+2ik_{s1}y} \\ +A_{p1}^*A_{p2}e^{-2ik_{s1}y} + A_{p2}^*A_{p1}e^{+2ik_{s1}y} \end{bmatrix} \right\}\hat{\mathbf{y}} \\ \left\{ -\frac{k_{s1}}{k}\frac{k_{p1}}{k}\operatorname{Im}\begin{bmatrix} +A_{s1}^*A_{s2}e^{-2ik_{s1}y} - A_{s2}^*A_{s1}e^{+2ik_{s1}y} \\ +A_{p1}^*A_{p2}e^{-2ik_{s1}y} - A_{p2}^*A_{p1}e^{+2ik_{s1}y} \end{bmatrix} \right\}\hat{\mathbf{z}} \end{pmatrix}, \quad (S36)$$

respectively. The discrepancy between the SAM and the vorticity of kinetic momentum is

$$\mathbf{S} - \frac{1}{2k^2}\nabla\times\mathbf{\Pi} = \hbar\sigma_1\hat{\mathbf{k}}_1 + \hbar\sigma_2\hat{\mathbf{k}}_2 + \hbar\sigma_c\left[\frac{W_c}{\hbar\omega}\frac{k_{p1}}{k}\hat{\mathbf{x}}\right].$$ (S37)

with the crossed helicity

$$\sigma_c = \frac{\text{Im}\left\{\left(A_{s1}^*A_{p2}e^{-2ik_{s1}y} - A_{s1}A_{p2}^*e^{2ik_{s1}y}\right) + \left(A_{s2}^*A_{p1}e^{2ik_{s1}y} - A_{s2}A_{p1}^*e^{-2ik_{s1}y}\right)\right\}}{\left(A_{p1}^*A_{p2}e^{-2ik_{s1}y} + A_{p2}^*A_{p1}e^{2ik_{s1}y}\right) + \left(A_{s1}^*A_{s2}e^{-2ik_{s1}y} + A_{s2}^*A_{s1}e^{2ik_{s1}y}\right)}.$$ (S38)

Here, the 1st/2nd term of numerator in the Eq. (S38) is the polarization ellipticity originated from the contribution between the $y/x$ component of wave 1 and $x/y$ component of wave 2 and the denominator is the crossed energy density in Eq. (S34).

On the other hand, considering the interference field, the directional vector of mean wavevector is

$$\hat{\mathbf{k}} = \frac{W}{\hbar\omega}\left(\frac{k_{p1}}{k}\hat{\mathbf{x}} + \frac{k_{s1}}{k}\frac{\left(A_{s1}^*A_{s1} + A_{p1}^*A_{p1} - A_{s2}^*A_{s2} - A_{p2}^*A_{p2}\right)}{W}\hat{\mathbf{y}} + 0\hat{\mathbf{z}}\right)$$
$$= \underbrace{\frac{W_1}{\hbar\omega}\left(\frac{k_{p1}}{k}\hat{\mathbf{x}} + \frac{k_{s1}}{k}\hat{\mathbf{y}} + 0\hat{\mathbf{z}}\right)}_{\hat{\mathbf{k}}_1} + \underbrace{\frac{W_2}{\hbar\omega}\left(\frac{k_{p1}}{k}\hat{\mathbf{x}} - \frac{k_{s1}}{k}\hat{\mathbf{y}} + 0\hat{\mathbf{z}}\right)}_{\hat{\mathbf{k}}_2} + \underbrace{\frac{W_c}{\hbar\omega}\frac{k_{p1}}{k}\hat{\mathbf{x}}}_{\hat{\mathbf{k}}_c} = \hat{\mathbf{k}}_1 + \hat{\mathbf{k}}_2 + \hat{\mathbf{k}}_c.$$ (S39)

By checking Eqs. (S37) and (S39), it can be found that the crossed energy density $W_c$ can be considered as propagating along the $x$-direction. In addition, the integral of this crossed spin component on the whole transverse plane ($yz$-plane) vanishes, which indicates that the this crossed term is local and does not affect the conservation of total angular momentum. Thus, we can rewrite the Eq. (S37) as

$$\mathbf{S} - \frac{1}{2k^2}\nabla\times\mathbf{\Pi} = \hbar\sigma_1\hat{\mathbf{k}}_1 + \hbar\sigma_2\hat{\mathbf{k}}_2 + \hbar\sigma_c\hat{\mathbf{k}}_c,$$ (S40)

with

$$\hat{\mathbf{k}}_c = \frac{W_c}{\hbar\omega}\frac{k_{p1}}{k}\hat{\mathbf{x}}.$$ (S41)

Interestingly, for a special case that $A_{s1}^* = A_{s2}$ and $A_{p1}^* = A_{p2}$, the two interfering plane waves are orthogonal and thus the crossed helicity vanishes. The Eq. (S40) can be rewritten as

$$\mathbf{S}_l = \mathbf{S} - \frac{1}{2k^2}\nabla\times\mathbf{\Pi} = \hbar\sigma_1\hat{\mathbf{k}}_1 + \hbar\sigma_2\hat{\mathbf{k}}_2.$$ (S42)

Therefore, we can understand the L-spin of interference field by the vector analysis. Moreover, the transverse spin can be re-expressed as

$$\mathbf{S}_t = \frac{\varepsilon e^{-2\kappa z}}{2\omega}\left(\begin{array}{c}0\hat{\mathbf{x}}\\ \left\{-\frac{\kappa^2}{k^2}\frac{k_{s1}}{k}\text{Im}\begin{bmatrix}+A_{s1}^*A_{p1} - A_{s1}A_{p1}^*\\ -A_{s2}^*A_{p2} + A_{s2}A_{p2}^*\end{bmatrix}\\ -\frac{\kappa}{k}\frac{k_{p1}}{k}\begin{bmatrix}+A_{s1}^*A_{s1} + A_{p1}^*A_{p1} + A_{s2}^*A_{s2} + A_{p2}^*A_{p2}\\ +A_{s1}^*A_{s2}e^{-2ik_{s1}y} + A_{s2}^*A_{s1}e^{+2ik_{s1}y} + A_{p1}^*A_{p2}e^{-2ik_{s1}y} + A_{p2}^*A_{p1}e^{+2ik_{s1}y}\end{bmatrix}\right\}\hat{\mathbf{y}}\\ \left\{-\frac{k_{s1}}{k}\frac{k_{p1}}{k}\text{Im}\begin{bmatrix}+A_{s1}^*A_{s2}e^{-2ik_{s1}y} - A_{s2}^*A_{s1}e^{+2ik_{s1}y}\\ +A_{p1}^*A_{p2}e^{-2ik_{s1}y} - A_{p2}^*A_{p1}e^{+2ik_{s1}y}\end{bmatrix}\right\}\hat{\mathbf{z}}\end{array}\right),$$ (S43)

and the directional vector of mean wavevector associated with the propagating direction is

$$\hat{\mathbf{k}} = \frac{W}{\hbar\omega}\left(\frac{k_{p1}}{k}\hat{\mathbf{x}} + 0\hat{\mathbf{y}} + 0\hat{\mathbf{z}}\right). \tag{S44}$$

It can be observed that the mean wavevector only contains $x$-component.

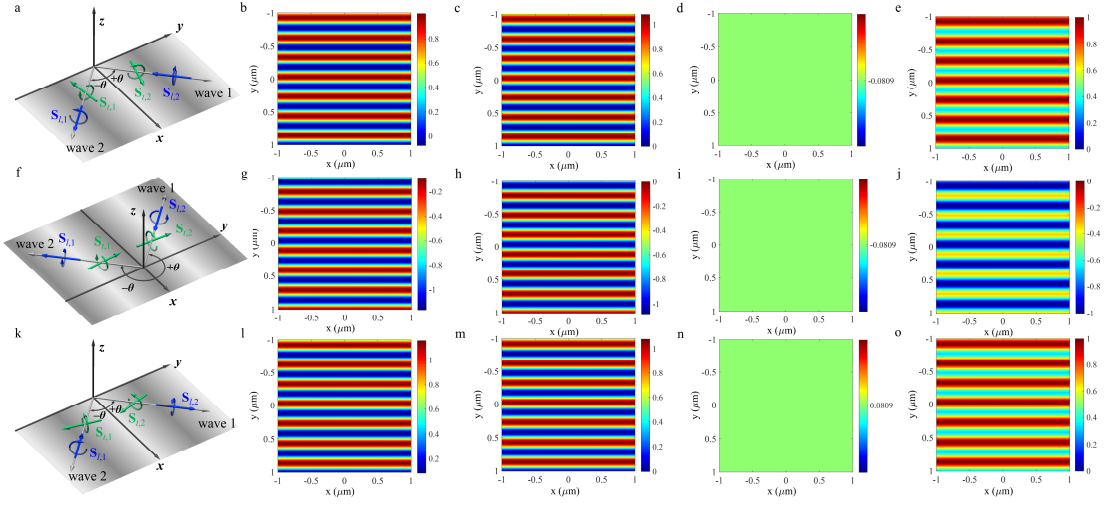

**Fig. S5.** Momentum properties of interference fields between two elliptically polarized evanescent plane waves. (a) Schematic diagram of the interference between two evanescent plane waves carrying opposite helices in the $xy$-plane. (b) The spatial distributions of the $x$-component kinetic momentum, when $\theta=45°$, $A_{s1} = 5+2i$ and $A_{s2} = 5-2i$. (c-d) The extracted helicity-irrelevant and helicity-related $x$-component kinetic momentum. (e) The $x$-component canonical momentum. (f-j) Same to (a-e) but with $\theta=135°$. (k-o) Same to (a-e) but with the opposite helicities, i.e., $A_{s1} = 5-2i$ and $A_{s2} = 5+2i$. In the calculation, $A_{p1} = A_{p2} =1$, $k_p = 1.5k$ and the wavelength of the waves is 632.8nm.

From the Eq. (S43), the $x$-component T-spin is always zero in the case, whereas the $y$-component T-spin contains two parts: the first part that is proportion to

$$\text{Im}\left[\left\{A_{s1}^* A_{p1} - A_{s1} A_{p1}^*\right\} - \left\{A_{s2}^* A_{p2} - A_{s2} A_{p2}^*\right\}\right]$$

is related to the helices of interfering plane waves; the second part that is originated from the exponential decaying of kinetic momentum (the helicity-irrelevant part of kinetic momentum includes the energy density as shown in Eq. (S26)) in the $z$-direction is helicity-independent. These make the total $y$-component SAM contains three parts: the L-spin, the helicity-dependent T-spin and the helicity-independent T-spin. In addition, the $z$-component that is proportional to the $y$-gradient of the helicity-irrelevant part of kinetic momentum is helicity-independent. To detailly exhibit the properties of EM spins, we give three examples here.

Firstly, as shown in Fig. S5(**a**), assuming $A_{p1} = A_{p2} =1$, $A_{s1} = 5+2i$, $A_{s2} = 5-2i$, and the propagating angle $\theta$ of the two-waves are +45º and –45º, respectively, the canonical momentum associated with the mean wavevector is along the +$x$-direction and varies periodically in the $y$-direction (Fig. S5(**e**)), while the kinetic momentum (Fig. S5(**b**)) has two components: the helicity-irrelevant component along the +$x$-direction and varying periodically in the $y$-direction (Fig. S5(**c**)), and the helicity-related component along the –$x$-direction and homogeneous in the $xy$-plane (Fig. S5(**d**)). Secondly, assuming the propagating angle $\theta$ of the two-waves are +135º and –135º (Fig. S5(**f**)), respectively, the canonical momentum associated with the propagating direction is inverted exactly (Fig. S5(**j**)). However, the variation of the colorbar values in Fig. 5(**b**) and Fig. 5(**g**) indicates the kinetic momentum is not opposite exactly. This is because only the helicity-irrelevant kinetic momentum is inverted (Fig. S5(**h**)) while the helicity-related kinetic momentum remains unchanged (Fig. S5(**i**)) due to the invariant of helical property in the field. Thirdly, assuming that the helical property of field is inverted (Fig. S5(**k**)) by setting $A_{s1}$ = 5–2i and $A_{s2}$ = 5+2i, the canonical momentum in Fig. S5(**o**) is same with that of Fig. S5(**e**) except for a translation in the *y*-direction (here, we translate all the physical quantities for the sake of comparison). In the case, the helicity-irrelevant kinetic momentum remains unchanged (Fig. S5(**m**)) and the helicity-related kinetic momentum is inverted (Fig. S5(**n**)). These make the colorbar values in Fig. 5(**l**) are not consistent with those in Fig. 5(**b**).

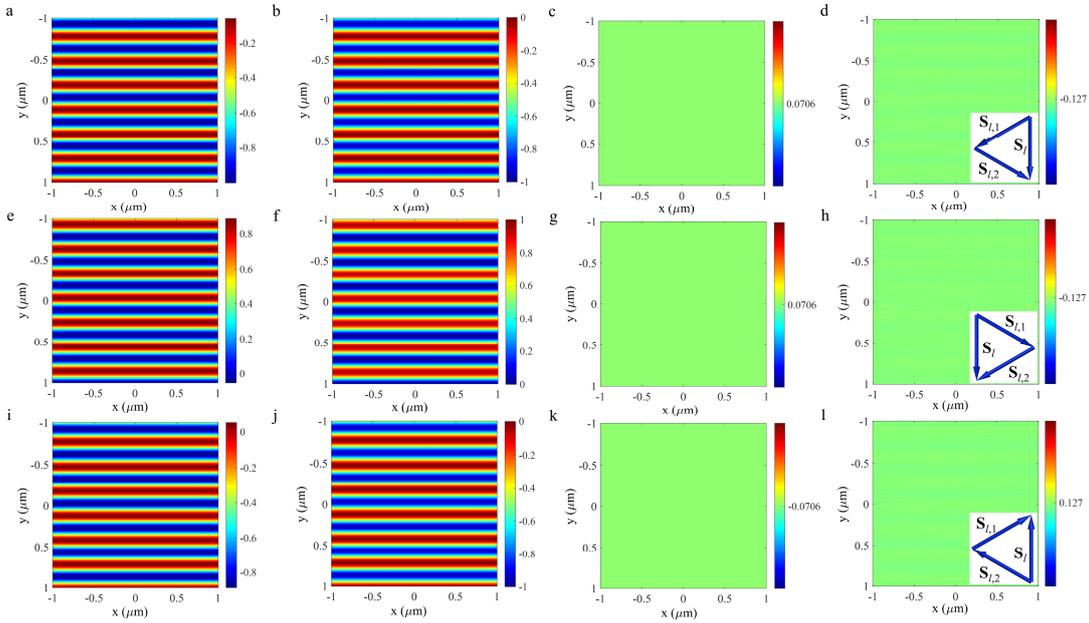

**Fig. S6.** Spin properties of interference fields between two elliptically polarized evanescent plane waves. (a) The spatial distributions of the *y*-component total SAM density, when $\theta$=45°, $A_{s1}$ = 5+2i and $A_{s2}$ = 5–2i. (b-c) The extracted helicity-independent and helicity-dependent *y*-component T-spin. (d) The extracted L-spin. (e-h) Same to (a-d) but with $\theta$=135°. (i-l) Same to (a-d) but with the opposite helicities, i.e., $A_{s1}$ = 5–2i and $A_{s2}$ = 5+2i. In the calculation, $A_{p1}$ = $A_{p2}$ =1, $k_p$ = 1.5$k$ and the wavelength of the waves is 632.8nm.

Subsequently, we investigate the spin properties of these three situations. The *z*-component SAMs are helicity-independent T-spin and possess the spin-momentum locking property, as shown in Figs. 3(**b, g, l**) in the main text. The *y*-component SAMs (Figs. S6(**a, e, i**)) contain three parts: the helicity-independent T-spins (Figs. S6(**b, f, j**)) are locked with the helicity-independent kinetic momenta; the helicity-dependent T-spins (Figs. S6(**c, g, k**)) possess the spin-momentum property and are locked with the helicity-dependent kinetic momenta; the L-spins (Figs. S6(**d, h, l**)) are helicity-dependent solely and do not possess the spin-momentum property.

In summary, the spin properties of interferential field between the elliptically polarized evanescent plane waves can be expressed as:

$$\mathbf{S}_l = \sum_i \hbar\sigma_i \hat{\mathbf{k}}_i + \sum_{i\neq j} \hbar\sigma_{ij}\hat{\mathbf{k}}_{ij}, \quad \mathbf{S}_t = \frac{1}{2k^2}\nabla\times\mathbf{\Pi} \quad \text{and} \quad \mathbf{S} = \mathbf{S}_l + \mathbf{S}_t. \tag{S45}$$

**Supplementary Section 5:** Spin properties of interfering propagating plane waves

The former section considers the spin properties of two-waves interference between two evanescent plane waves. In the section, we will generalize the analysis to the two-waves interference in free space. We first consider the interference of two propagating plane waves (wave 1: $A_{s1}$, $A_{p1}$ and wave 2: $A_{s2}$, $A_{p2}$) propagating along the $xz$-plane with an angle $2\Theta_1$ between their wave vectors

$$\mathbf{k}_{1,2} = k_{p1}\hat{\mathbf{p}} \pm k_{z1}\hat{\mathbf{z}},$$

where $k_{p1}=k\cos\Theta_1$ and $k_{z1}=k\sin\Theta_1$ as shown in Fig. S7(**a**). For each individual plane wave, the energy density, direction of local wave vector, and SAM can be expressed as:

$$W_1 = \frac{\varepsilon}{2}\{A_{s1}^*A_{s1}+A_{p1}^*A_{p1}\} \quad W_2 = \frac{\varepsilon}{2}\{A_{s2}^*A_{s2}+A_{p2}^*A_{p2}\}, \tag{S46}$$

$$\hat{\mathbf{k}}_1 = \frac{W_1}{\hbar\omega}\left(k_{p1}/k\,\hat{\mathbf{x}} + 0\hat{\mathbf{y}} + k_{z1}/k\,\hat{\mathbf{z}}\right) \quad \hat{\mathbf{k}}_2 = \frac{W_2}{\hbar\omega}\left(k_{p1}/k\,\hat{\mathbf{x}} + 0\hat{\mathbf{y}} - k_{z1}/k\,\hat{\mathbf{z}}\right), \tag{S47}$$

and

$$\mathbf{S}_{l,1} = \hbar\frac{\text{Im}\{A_{s1}^*A_{p1}-A_{p1}^*A_{s1}\}}{A_{s1}^*A_{s1}+A_{p1}^*A_{p1}}\hat{\mathbf{k}}_1 = \hbar\sigma_1\hat{\mathbf{k}}_1 \quad \mathbf{S}_{l,2} = \hbar\frac{\text{Im}\{A_{s2}^*A_{p2}-A_{p2}^*A_{s2}\}}{A_{s2}^*A_{s2}+A_{p2}^*A_{p2}}\hat{\mathbf{k}}_2 = \hbar\sigma_2\hat{\mathbf{k}}_2. \tag{S48}$$

These results match well with those of Equations (S9), (S12) and (S13).

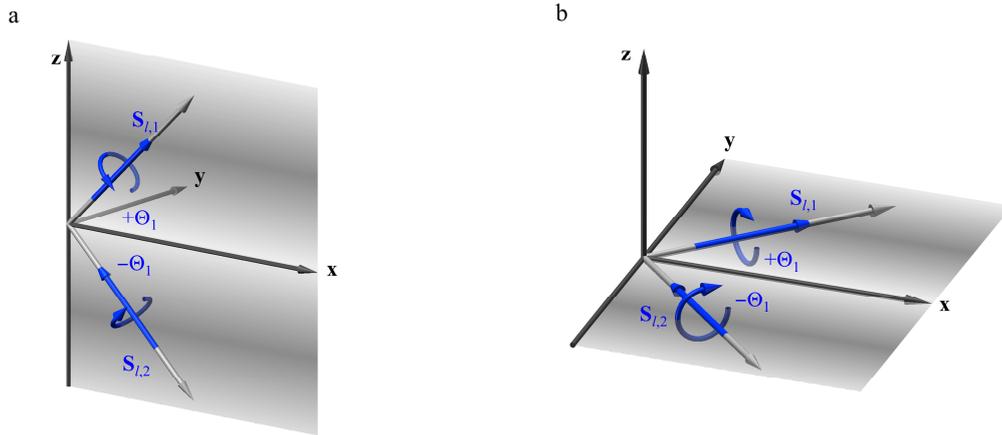

**Fig. S7.** (a) The schematic diagram of the interference between two elliptically polarized propagating plane waves in $xz$-plane. The intersected angle between the individual wave and the $x$-axis is $\Theta_1$. (b) The schematic diagram of the interference between two elliptically polarized propagating plane waves in $xy$-plane. The intersected angle between the individual wave and the $x$-axis is $\Theta_1$.

For the interference field, the energy density is

$$W = \frac{\varepsilon}{2} \left\{ \underbrace{A_{s1}^* A_{s1} + A_{p1}^* A_{p1}}_{W_1} + \underbrace{A_{s2}^* A_{s2} + A_{p2}^* A_{p2}}_{W_2} + \underbrace{\frac{k_{p1}^2}{k^2} \left( A_{s1}^* A_{s2} e^{-2ik_{z1}z} + A_{s2}^* A_{s1} e^{2ik_{z1}z} + A_{p1}^* A_{p2} e^{-2ik_{z1}z} + A_{p2}^* A_{p1} e^{2ik_{z1}z} \right)}_{W_c} \right\} = W_1 + W_2 + W_c, \quad (S49)$$

The term $W_c$ represents the crossed energy density.

The SAM and the T-spin given by the vorticity of kinetic momentum are

$$\mathbf{S} = \frac{\varepsilon}{2\omega} \left\{ \begin{array}{l} \frac{k_{p1}}{k} \text{Im} \left[ \begin{array}{l} +\left( A_{s1}^* A_{p1} - A_{s1} A_{p1}^* \right) + \left( A_{s2}^* A_{p2} - A_{s2} A_{p2}^* \right) \\ +\left( A_{s1}^* A_{p2} e^{-2ik_{z1}z} - A_{s1} A_{p2}^* e^{2ik_{z1}z} \right) + \left( A_{s2}^* A_{p1} e^{2ik_{z1}z} - A_{s2} A_{p1}^* e^{-2ik_{z1}z} \right) \end{array} \right] \hat{\mathbf{x}} \\ \frac{k_{p1}}{k} \frac{k_{z1}}{k} \text{Im} \left[ +\left( A_{s1}^* A_{s2} e^{-2ik_{z1}z} - A_{s1} A_{s2}^* e^{2ik_{z1}z} \right) + \left( A_{p1}^* A_{p2} e^{-2ik_{z1}z} - A_{p1} A_{p2}^* e^{2ik_{z1}z} \right) \right] \hat{\mathbf{y}} \\ \frac{k_{z1}}{k} \text{Im} \left[ +\left( A_{s1}^* A_{p1} - A_{s1} A_{p1}^* \right) - \left( A_{s2}^* A_{p2} - A_{s2} A_{p2}^* \right) \right] \hat{\mathbf{z}} \end{array} \right\} \quad (S50)$$

and

$$\mathbf{S}_t = \frac{1}{2k^2} \nabla \times \mathbf{\Pi}$$
$$= \frac{\varepsilon}{2\omega} \text{Im} \left( \begin{array}{l} \frac{k_{p1} k_{z1}^2}{k^3} \left[ +\left( A_{s1}^* A_{p2} e^{-2ik_{z1}z} - A_{s1} A_{p2}^* e^{2ik_{z1}z} \right) + \left( A_{s2}^* A_{p1} e^{2ik_{z1}z} - A_{s2} A_{p1}^* e^{-2ik_{z1}z} \right) \right] \hat{\mathbf{x}} \\ \frac{k_{p1} k_{z1}}{k^2} \left[ +\left( A_{s1}^* A_{s2} e^{-2ik_{z1}z} - A_{s2}^* A_{s1} e^{2ik_{z1}z} \right) + \left( A_{p1}^* A_{p2} e^{-2ik_{z1}z} - A_{p2}^* A_{p1} e^{2ik_{z1}z} \right) \right] \hat{\mathbf{y}} \\ 0 \hat{\mathbf{z}} \end{array} \right), \quad (S51)$$

respectively. The discrepancy between the SAM and the vorticity of kinetic momentum is

$$\mathbf{S} - \frac{1}{2k^2} \nabla \times \mathbf{\Pi} = \hbar \sigma_1 \hat{\mathbf{k}}_1 + \hbar \sigma_2 \hat{\mathbf{k}}_2 + \hbar \sigma_c \left[ \frac{W_c(\mathbf{r})}{\hbar \omega} \frac{k_{p1}}{k} \hat{\mathbf{x}} \right]. \quad (S52)$$

with the crossed helicity

$$\sigma_c = \frac{\text{Im}\left\{ \left( A_{s1}^* A_{p2} e^{-2ik_{z1}z} - A_{s1} A_{p2}^* e^{2ik_{z1}z} \right) + \left( A_{s2}^* A_{p1} e^{2ik_{z1}z} - A_{s2} A_{p1}^* e^{-2ik_{z1}z} \right) \right\}}{\left( A_{p1}^* A_{p2} e^{-2ik_{z1}z} + A_{p2}^* A_{p1} e^{2ik_{z1}z} \right) + \left( A_{s1}^* A_{s2} e^{-2ik_{z1}z} + A_{s2}^* A_{s1} e^{2ik_{z1}z} \right)}. \quad (S53)$$

Here, the 1st/2nd term of numerator in the Eq. (S53) represents the polarization ellipticity between the *s/p* component of wave 1 and *p/s* component of wave 2, and the denominator of $\sigma_c$ is the crossed energy density Eq. (S49).

On the other hand, considering the interference field, the directional vector of mean wavevector is

$$\hat{\mathbf{k}} = \frac{W}{\hbar \omega} \left( +\frac{k_{p1}}{k} \hat{\mathbf{x}} + 0\hat{\mathbf{y}} + \frac{k_{z1}}{k} \frac{\varepsilon}{2} \frac{\left( A_{p1}^* A_{p1} + A_{s1}^* A_{s1} - A_{p2}^* A_{p2} - A_{s2}^* A_{s2} \right)}{W} \hat{\mathbf{z}} \right)$$
$$= \frac{W_1}{\hbar \omega} \left( +\frac{k_{p1}}{k} \hat{\mathbf{x}} + 0\hat{\mathbf{y}} + \frac{k_{z1}}{k} \hat{\mathbf{z}} \right) + \frac{W_2}{\hbar \omega} \left( +\frac{k_{p1}}{k} \hat{\mathbf{x}} + 0\hat{\mathbf{y}} - \frac{k_{z1}}{k} \hat{\mathbf{z}} \right) + \frac{W_c}{\hbar \omega} \frac{k_{p1}}{k} \hat{\mathbf{x}} = \hat{\mathbf{k}}_1 + \hat{\mathbf{k}}_2 + \hat{\mathbf{k}}_c \quad (S54)$$

Thus, we can rewrite the Equation (S52) as

$$\mathbf{S} - \frac{1}{2k^2}\nabla \times \mathbf{\Pi} = \hbar\sigma_1\hat{\mathbf{k}}_1 + \hbar\sigma_2\hat{\mathbf{k}}_2 + \hbar\sigma_c\hat{\mathbf{k}}_c, \quad (S55)$$

with

$$\hat{\mathbf{k}}_c = \frac{W_c}{\hbar\omega}\frac{k_{p1}}{k}\hat{\mathbf{x}}. \quad (S56)$$

For a special case that $A_{s1}^* = A_{s2}$ and $A_{p1}^* = A_{p2}$, the two interfering plane waves are orthogonal and thus the crossed helicity $\sigma_c$ vanishes. The Eq. (S55) can be rewritten as

$$\mathbf{S} - \frac{1}{2k^2}\nabla \times \mathbf{\Pi} = \hbar\sigma_1\hat{\mathbf{k}}_1 + \hbar\sigma_2\hat{\mathbf{k}}_2. \quad (S57)$$

Equations (S55) and (S57) shows that the difference between the total SAM and the T-spin given by the vorticity of kinetic momentum is the L-spin.

Subsequently, as shown in Fig. S7(**b**), we consider a superposition of two propagating plane waves (wave 1: $A_{s1}$, $A_{p1}$ and wave 2: $A_{s2}$, $A_{p2}$) propagating along $xy$-plane with an angle $2\Theta_1$ between their wave vectors

$$\mathbf{k}_{1,2} = k_{p1}\hat{\mathbf{x}} \pm k_{s1}\hat{\mathbf{y}}$$

where $k_{p1}=k\cos\Theta_1$ and $k_{s1}=k\sin\Theta_1$. The derivations are similar to the former interference in the $xz$-plane, we summarize the results briefly.

For the interference field, the energy density is

$$W = \frac{\varepsilon}{2}\left\{\underbrace{A_{p1}^*A_{p1} + A_{s1}^*A_{s1}}_{W_1} + \underbrace{A_{p2}^*A_{p2} + A_{s2}^*A_{s2}}_{W_2} + \underbrace{\frac{k_{p1}^2}{k^2}\left(A_{p2}^*A_{p1}e^{2ik_{s1}y} + A_{p1}^*A_{p2}e^{-2ik_{s1}y} + A_{s2}^*A_{s1}e^{2ik_{s1}y} + A_{s1}^*A_{s2}e^{-2ik_{s1}y}\right)}_{W_c}\right\} = W_1 + W_2 + W_c, \quad (S59)$$

with the crossed energy term $W_c$.

The SAM and the T-spin given by the vorticity of kinetic momentum are

$$\mathbf{S} = \frac{\varepsilon}{2\omega}\text{Im}\left\{\begin{array}{l}\frac{k_{p1}}{k}\left[+\left(A_{s1}^*A_{p1} - A_{s1}A_{p1}^*\right) + \left(A_{s2}^*A_{p2} - A_{s2}A_{p2}^*\right)\right.\\ \left.+\left(A_{s1}^*A_{p2}e^{-2ik_{s1}y} - A_{s1}A_{p2}^*e^{2ik_{s1}y}\right) + \left(A_{s2}^*A_{p1}e^{2ik_{s1}y} - A_{s2}A_{p1}^*e^{-2ik_{s1}y}\right)\right]\hat{\mathbf{x}}\\ \frac{k_{s1}}{k}\left[+\left(A_{s1}^*A_{p1} - A_{s1}A_{p1}^*\right) - \left(A_{s2}^*A_{p2} - A_{s2}A_{p2}^*\right)\right]\hat{\mathbf{y}}\\ \frac{k_{s1}k_{p1}}{k^2}\left[-\left(A_{s1}^*A_{s2}e^{-2ik_{s1}y} - A_{s1}A_{s2}^*e^{2ik_{s1}y}\right) - \left(A_{p1}^*A_{p2}e^{-2ik_{s1}y} - A_{p1}A_{p2}^*e^{2ik_{s1}y}\right)\right]\hat{\mathbf{z}}\end{array}\right\}, \quad (S60)$$

and

$$\mathbf{S}_t = \frac{1}{2k^2}\nabla \times \mathbf{\Pi} = \frac{\varepsilon}{2\omega}\text{Im}\left(\begin{array}{c}\frac{k_{p1}k_{s1}^2}{k^3}\left[A_{s1}^*A_{p2}e^{-2ik_{s1}y} - A_{s1}A_{p2}^*e^{2ik_{s1}y} + A_{s2}^*A_{p1}e^{2ik_{s1}y} - A_{s2}A_{p1}^*e^{-2ik_{s1}y}\right]\hat{\mathbf{x}}\\ 0\hat{\mathbf{y}}\\ \frac{k_{p1}k_{s1}}{k^2}\left[-A_{s1}^*A_{s2}e^{-2ik_{s1}y} + A_{s1}A_{s2}^*e^{2ik_{s1}y} - A_{p1}^*A_{p2}e^{-2ik_{s1}y} + A_{p1}A_{p2}^*e^{2ik_{s1}y}\right]\hat{\mathbf{z}}\end{array}\right), \quad (S61)$$

respectively. The discrepancy between the SAM and the vorticity of kinetic momentum is

$$\mathbf{S} - \frac{1}{2k^2}\nabla \times \mathbf{\Pi} = \hbar\sigma_1\hat{\mathbf{k}}_1 + \hbar\sigma_2\hat{\mathbf{k}}_2 + \hbar\sigma_c\left[\frac{W_c}{\hbar\omega}\frac{k_{p1}}{k}\hat{\mathbf{x}}\right]. \quad (S62)$$

with the crossed helicity

$$\sigma_c = \frac{\mathrm{Im}\left\{\left(A_{s1}^* A_{p2} e^{-2ik_{s1}y} - A_{s1} A_{p2}^* e^{2ik_{s1}y}\right) + \left(A_{s2}^* A_{p1} e^{2ik_{s1}y} - A_{s2} A_{p1}^* e^{-2ik_{s1}y}\right)\right\}}{A_{p2}^* A_{p1} e^{2ik_{s1}y} + A_{p1}^* A_{p2} e^{-2ik_{s1}y} + A_{s2}^* A_{s1} e^{2ik_{s1}y} + A_{s1}^* A_{s2} e^{-2ik_{s1}y}}. \tag{S63}$$

The directional vector of mean wavevector of the interference field is

$$\hat{\mathbf{k}} = \frac{W_1}{\hbar\omega}\left(\frac{k_{p1}}{k}\hat{\mathbf{x}} + \frac{k_{s1}}{k}\hat{\mathbf{y}} + 0\hat{\mathbf{z}}\right) + \frac{W_2}{\hbar\omega}\left(\frac{k_{p1}}{k}\hat{\mathbf{x}} - \frac{k_{s1}}{k}\hat{\mathbf{y}} + 0\hat{\mathbf{z}}\right) + \frac{W_c}{\hbar\omega}\frac{k_{p1}}{k}\hat{\mathbf{x}} = \hat{\mathbf{k}}_1 + \hat{\mathbf{k}}_2 + \hat{\mathbf{k}}_c. \tag{S64}$$

The Eq. (S62) can be rewritten as

$$\mathbf{S} - \frac{1}{2k^2}\nabla\times\mathbf{\Pi} = \hbar\sigma_1\hat{\mathbf{k}}_1 + \hbar\sigma_2\hat{\mathbf{k}}_2 + \hbar\sigma_c\hat{\mathbf{k}}_c, \tag{S65}$$

with directional vector

$$\hat{\mathbf{k}}_c = \frac{W_c}{\hbar\omega}\frac{k_{p1}}{k}\hat{\mathbf{x}}. \tag{S66}$$

In the case that $A_{s1}^* = A_{s2}$ and $A_{p1}^* = A_{p2}$, the crossed helicity $\sigma_c$ vanishes and the Eq. (S65) can be rewritten as

$$\mathbf{S} - \frac{1}{2k^2}\nabla\times\mathbf{\Pi} = \hbar\sigma_1\hat{\mathbf{k}}_1 + \hbar\sigma_2\hat{\mathbf{k}}_2. \tag{S67}$$

Equations (S65) and (S67) also shows that the difference between the total SAM and the T-spin given by the vorticity of kinetic momentum is the L-spin.

In summary, the spin-momentum properties of interference field of the elliptically polarized propagating plane waves can be summed as:

$$\mathbf{S}_l = \sum_i \hbar\sigma_i\hat{\mathbf{k}}_i + \sum_{i\neq j}\hbar\sigma_{ij}\hat{\mathbf{k}}_{ij}, \quad \mathbf{S}_t = \frac{1}{2k^2}\nabla\times\mathbf{\Pi} \quad \text{and} \quad \mathbf{S} = \mathbf{S}_l + \mathbf{S}_t. \tag{S68}$$

**Supplementary Section 6:** Spin-momentum equations for the generic electromagnetic field

In the former sections, we demonstrate that the T-spin is originated from the transverse inhomogeneities of EM field, while the L-spin is determined by the EM helicity solely. By employing the equations (S5-S8), we can formulate a novel set of Maxwell-like spin-momentum equations in the uniform nondispersive space as:

$$\nabla\cdot\mathbf{\Pi} = 0, \tag{S69}$$

$$\nabla\cdot\mathbf{S} = 0, \tag{S70}$$

$$\nabla\times\mathbf{\Pi} = 2k^2(\mathbf{S}-\mathbf{S}_l) = 2k^2\mathbf{S}_t, \tag{S71}$$

$$\nabla\times\mathbf{S} = 2(\mathbf{\Pi}-\mathbf{P}), \tag{S72}$$

and the Helmholtz-like equation as

$$\nabla^2\mathbf{S} + 4k^2(\mathbf{S}-\mathbf{S}_l) = 2\nabla\times\mathbf{P}. \tag{S73}$$

With these equations, one can investigate the spin-orbit coupling of EM field and control the spin-orbit interaction actively (10, 28, 30).

**Supplementary Section 7:** Transverse spin and Berry curvature

For a scalar wave in free space or a single polarized wave at an optical interface (surface plasmons polaritons (S4) or Bloch surface wave (S5)), the kinetic momentum can be expressed as the flux density of optical potential (54). We take the surface plasmons polaritons for example. The kinetic momentum can be expressed as

$$\mathbf{\Pi} = \frac{\varepsilon k^2 k_p^2}{4\omega} \text{Im}\{\Psi^* \nabla \Psi - \Psi \nabla \Psi^*\} = -\frac{\varepsilon k^2 k_p^2}{2\omega} \langle \Psi | i\nabla | \Psi \rangle. \tag{S74}$$

Here, $\Psi$ indicates the Hertz potential for transverse magnetic surface modes. Thus, from the Eq. (S30), the T-spin can be calculated as

$$\mathbf{S}_t = \frac{1}{2k^2} \nabla \times \mathbf{\Pi} = -\frac{\varepsilon k_p^2}{4\omega} \langle \nabla \Psi | \times i | \nabla \Psi \rangle, \tag{S75}$$

which has a similar form with the Berry curvature of the Hertz potential.

Then, for the generic EM field, through the tedious derivations, one can obtain that

$$\mathbf{S}_t = \frac{1}{2k^2} \nabla \times \mathbf{\Pi} = \frac{\mathbf{S}}{2} - \frac{1}{8\omega^2} \text{Re} \begin{Bmatrix} -(\nabla \odot \mathbf{E}^*) \cdot \mathbf{H} - (\nabla \odot \mathbf{E})^T \cdot \mathbf{H}^* \\ +(\nabla \odot \mathbf{H}^*) \cdot \mathbf{E} + (\nabla \odot \mathbf{H})^T \cdot \mathbf{E}^* \end{Bmatrix}, \tag{S76}$$

where

$$\mathbf{r}_1 \odot \mathbf{r}_2 = \begin{pmatrix} x_i^1 x_i^2 & x_j^1 x_i^2 & x_k^1 x_i^2 \\ x_i^1 x_j^2 & x_j^1 x_j^2 & x_k^1 x_j^2 \\ x_i^1 x_k^2 & x_j^1 x_k^2 & x_k^1 x_k^2 \end{pmatrix}. \tag{S77}$$

Note that the second part in the right-hand side of Eq. (S86) has a same structure as the quantum 2-form (55) that generates the Berry phase associated with a circuit, which indicates a spin-orbit interaction in the optical system. Thus, we can conclude that the T-spin is closely related to the Berry curvature and the evolution of geometric phase in the EM systems.

**Supplementary Section 8:** Spin properties of focusing circular polarization light in free space

To explain the relationship between the T-spin and the geometric phase, we exhibit the spin properties of focusing circular polarization light. Assuming that the circularly polarized light propagates along $z$-direction, the electric field of incident light in cylindrical coordinate ($\rho$, $\varphi$, $z$) systems can be expressed as:

$$\mathbf{E} = \frac{1}{\sqrt{2}} (\hat{\mathbf{e}}_x \pm \sigma i \hat{\mathbf{e}}_y) e^{ikz} = \frac{1}{\sqrt{2}} (e^{\sigma i\varphi} \hat{\mathbf{e}}_\rho + \sigma i e^{\sigma i\varphi} \hat{\mathbf{e}}_\varphi) e^{ikz} = (E_\rho \hat{\mathbf{e}}_\rho + E_\varphi \hat{\mathbf{e}}_\varphi) e^{ikz}. \tag{S78}$$

Here, $\sigma = \pm 1$ denote the right-handed and left-handed circularly polarized light. From the Richard-Wolf vectorial diffraction theory (S6), the focused electric and magnetic field components can be given by

$$\mathbf{E}_f = \frac{ikfe^{-ikf}}{2\pi} \int_0^{\theta_{max}} \int_0^{2\pi} \sqrt{\frac{n_i \cos\theta}{n_o}} A(\theta) \begin{pmatrix} \begin{bmatrix} +t_p \cos\theta \cos\varphi E_\rho \\ -t_s \sin\varphi E_\varphi \end{bmatrix} \hat{\mathbf{e}}_x \\ \begin{bmatrix} +t_p \cos\theta \sin\varphi E_\rho \\ +t_s \cos\varphi E_\varphi \end{bmatrix} \hat{\mathbf{e}}_y \\ \begin{bmatrix} -t_p \sin\theta E_\rho \end{bmatrix} \hat{\mathbf{e}}_z \end{pmatrix} \cdot e^{ik[z\cos\theta + r\sin\theta\cos(\varphi-\phi)]} \sin\theta d\varphi d\theta \quad (S79)$$

and

$$\mathbf{H}_f = \frac{ikfe^{-ikf}}{2\pi\eta} \int_0^{\theta_{max}} \int_0^{2\pi} \sqrt{\frac{n_i \cos\theta}{n_o}} A(\theta) \begin{pmatrix} \begin{bmatrix} -t_p \sin\varphi E_\rho \\ -t_s \cos\theta \cos\varphi E_\varphi \end{bmatrix} \hat{\mathbf{e}}_x \\ \begin{bmatrix} +t_p \cos\varphi E_\rho \\ -t_s \cos\theta \sin\varphi E_\varphi \end{bmatrix} \hat{\mathbf{e}}_y \\ \begin{bmatrix} +t_s \sin\theta E_\varphi \end{bmatrix} \hat{\mathbf{e}}_z \end{pmatrix} \cdot e^{ik[z\cos\theta + r\sin\theta\cos(\varphi-\phi)]} \sin\theta d\varphi d\theta. \quad (S80)$$

Here, $f$ is the focal length; $n_i$ and $n_o$ are the refractive indices of incident and focusing plane, respectively; $t_p(\theta)$ and $t_s(\theta)$ are the Fresnel transmission coefficients for $p$-polarization and $s$-polarization (we ignore the $\theta$-dependent in the expressions for simplicity.); $\theta=\mathrm{asin}(\rho/f)$ and $\theta_{max}=\mathrm{asin}(\mathrm{NA}/n_o)$ with NA denoting the numerical aperture of lens. In the focal plane ($z = 0$), the field components are respectively

$$\mathbf{E}_f = \int_0^{\theta_{max}} B(\theta) \begin{pmatrix} \{(t_p \cos\theta + t_s)J_0(kr\sin\theta) - (t_p\cos\theta - t_s)J_{2\sigma}(kr\sin\theta)\}e^{\sigma i\varphi}\hat{\mathbf{e}}_r \\ +\sigma i\{(t_p\cos\theta + t_s)J_0(kr\sin\theta) + (t_p\cos\theta - t_s)J_{2\sigma}(kr\sin\theta)\}e^{\sigma i\varphi}\hat{\mathbf{e}}_\phi \\ \{-2i^\sigma t_p \sin\theta J_\sigma(kr\sin\theta)\}e^{\sigma i\varphi}\hat{\mathbf{e}}_z \end{pmatrix} d\theta, \quad (S81)$$

$$\mathbf{H}_f = \int_0^{\theta_{max}} \frac{B(\theta)}{\eta} \begin{pmatrix} -\sigma i\{(t_p + t_s\cos\theta)J_0(kr\sin\theta) + (t_p - t_s\cos\theta)J_{2\sigma}(kr\sin\theta)\}e^{\sigma i\varphi}\hat{\mathbf{e}}_r \\ \{(t_p + t_s\cos\theta)J_0(kr\sin\theta) - (t_p - t_s\cos\theta)J_{2\sigma}(kr\sin\theta)\}e^{\sigma i\varphi}\hat{\mathbf{e}}_\phi \\ \{-2t_s \sin\theta J_\sigma(kr\sin\theta)\}e^{\sigma i\varphi}\hat{\mathbf{e}}_z \end{pmatrix} d\theta. \quad (S82)$$

Here, $B(\theta) = \frac{ikfe^{-ikf}}{2} A(\theta)\sin\theta \sqrt{\frac{n_i \cos\theta}{2n_o}}$ and $J_\sigma$ is the $\sigma$-order Bessel function of first kind. The detailed derivations are not given here due to the complexity of the expressions. The $z$-component SAM is

$$S_z = \hbar\sigma C(r) \quad (S83)$$

with

$$C(r) = \int_0^{\theta_{max}} \int_0^{\theta_{max}} \mathrm{Re}\left[\frac{\varepsilon B_1^* B_2}{2\hbar\omega}\right] \begin{Bmatrix} +(t_{s,1} + t_{p,1}\cos\theta_1)(t_{s,2} + t_{p,2}\cos\theta_2)J_0(kr\sin\theta_1)J_0(kr\sin\theta_2) \\ +(t_{p,1} + t_{s,1}\cos\theta_1)(t_{p,2} + t_{s,2}\cos\theta_2)J_0(kr\sin\theta_1)J_0(kr\sin\theta_2) \\ -(t_{s,1} - t_{p,1}\cos\theta_1)(t_{s,2} - t_{p,2}\cos\theta_2)J_2(kr\sin\theta_1)J_2(kr\sin\theta_2) \\ -(t_{p,1} - t_{s,1}\cos\theta_1)(t_{p,2} - t_{s,2}\cos\theta_2)J_2(kr\sin\theta_1)J_2(kr\sin\theta_2) \end{Bmatrix} d\theta_1 d\theta_2.$$

Here, the subscripts 1 and 2 represent the different integral quantities. Since the coefficient $C(r)$ is helicity-independent, it can be concluded that the $z$-component SAM is helicity-dependent.

On the other hand, the $z$-component of vorticity of kinetic momentum is

$$S_{t,z} = \left(\frac{1}{2k^2}\nabla\times\mathbf{\Pi}\right)_z = \hbar\sigma D(r) \tag{S84}$$

with

$$D(r) = \int_0^{\theta_{max}}\int_0^{\theta_{max}} \text{Re}\left[\frac{\varepsilon B_1^* B_2}{2\hbar\omega}\right] \begin{Bmatrix} +\sin\theta_2\sin\theta_2\left[t_{s,1}t_{s,2}+t_{p,1}t_{s,2}\cos\theta_1\right]J_0(kr\sin\theta_1)J_0(kr\sin\theta_2) \\ +\sin\theta_1\sin\theta_1\left[t_{p,1}t_{p,2}+t_{p,1}t_{s,2}\cos\theta_2\right]J_0(kr\sin\theta_1)J_0(kr\sin\theta_2) \\ -\sin\theta_1\sin\theta_2\left[t_{s,1}t_{s,2}+t_{p,1}t_{s,2}\cos\theta_1\right]J_1(kr\sin\theta_1)J_1(kr\sin\theta_2) \\ -\sin\theta_1\sin\theta_2\left[t_{p,1}t_{p,2}+t_{p,1}t_{s,2}\cos\theta_2\right]J_1(kr\sin\theta_1)J_1(kr\sin\theta_2) \\ +\sin\theta_1\sin\theta_1\left[t_{p,1}t_{p,2}-t_{p,1}t_{s,2}\cos\theta_2\right]J_0(kr\sin\theta_1)J_2(kr\sin\theta_2) \\ +\sin\theta_1\sin\theta_2\left[t_{p,1}t_{p,2}-t_{p,1}t_{s,2}\cos\theta_2\right]J_1(kr\sin\theta_1)J_2'(kr\sin\theta_2) \\ +\sin\theta_2\sin\theta_2\left[t_{s,1}t_{s,2}-t_{p,1}t_{s,2}\cos\theta_1\right]J_2(kr\sin\theta_1)J_0(kr\sin\theta_2) \\ +\sin\theta_1\sin\theta_2\left[t_{s,1}t_{s,2}-t_{p,1}t_{s,2}\cos\theta_1\right]J_2'(kr\sin\theta_1)J_1(kr\sin\theta_2) \end{Bmatrix} d\theta_1 d\theta_2 .$$

This T-spin is originated from the helicity-related kinetic momentum in the $xy$-plane. Obviously, $D(r)$ is non-zero, which indicate a helicity-dependent T-spin. Thus, we can conclude that the difference between the $z$-components of SAM and the vorticity of kinetic momentum is the helicity-dependent L-spin:

$$\mathbf{S}_{l,z} = \left(\mathbf{S}-\frac{1}{2k^2}\nabla\times\mathbf{\Pi}\right)_z = \hbar\sigma\left[C(r)-D(r)\right]. \tag{S85}$$

Subsequently, from the derivations, we find that the azimuthal T-spin is

$$\left(\frac{1}{2\omega^2}\nabla\times\mathbf{P}\right)_\varphi$$

$$= \int_0^{\theta_{max}}\int_0^{\theta_{max}} \text{Re}\left[\frac{\varepsilon B_1^* B_2}{2\omega}\right] \begin{Bmatrix} +\sin\theta_1\left[t_{p,1}\cos\theta_1+t_{s,1}\right]\left[t_{p,2}+t_{s,2}\cos\theta_2\right]J_1(kr\sin\theta_1)J_0(kr\sin\theta_2) \\ +\sin\theta_2\left[t_{p,1}\cos\theta_1+t_{s,1}\right]\left[t_{p,2}+t_{s,2}\cos\theta_2\right]J_0(kr\sin\theta_1)J_1(kr\sin\theta_2) \\ -\sin\theta_1\left[t_{p,1}\cos\theta_1-t_{s,1}\right]\left[t_{p,2}-t_{s,2}\cos\theta_2\right]J_2'(kr\sin\theta_1)J_2(kr\sin\theta_2) \\ -\sin\theta_2\left[t_{p,1}\cos\theta_1-t_{s,1}\right]\left[t_{p,2}-t_{s,2}\cos\theta_2\right]J_2(kr\sin\theta_1)J_2'(kr\sin\theta_2) \end{Bmatrix} d\theta_1 d\theta_2 .$$

(S86)

Equation (S86) indicates that the azimuthal component of T-spin is helicity-independent. This component is regarded as the helicity-independent T-spin and is perpendicular to the optical axis. It is worth noting that there is coupling term (given by the multiplying of $t_p$ and $t_s$) in the azimuthal SAM, and hence the distinction between the azimuthal SAM and azimuthal T-spin is

$$\left(\mathbf{S}-\frac{1}{2\omega^2}\nabla\times\mathbf{P}\right)_\varphi \approx$$

$$\int_0^{\theta_{max}}\int_0^{\theta_{max}}\text{Re}\left[\frac{\varepsilon B_1^* B_2}{2\omega}\right]\begin{Bmatrix}+\sin\theta_2\left[t_{p,1}t_{s,2}\left(1-\cos\theta_1\cos\theta_2\right)+t_{s,1}t_{s,2}\left(\cos\theta_1-\cos\theta_2\right)\right]J_0\left(kr\sin\theta_1\right)J_1\left(kr\sin\theta_2\right)\\+\sin\theta_1\left[t_{p,1}t_{s,2}\left(1-\cos\theta_1\cos\theta_2\right)-t_{p,1}t_{p,2}\left(\cos\theta_1-\cos\theta_2\right)\right]J_1\left(kr\sin\theta_1\right)J_0\left(kr\sin\theta_2\right)\\+\sin\theta_1\left[t_{p,1}t_{s,2}\left(1-\cos\theta_1\cos\theta_2\right)+t_{p,1}t_{p,2}\left(\cos\theta_1-\cos\theta_2\right)\right]J_1\left(kr\sin\theta_1\right)J_2\left(kr\sin\theta_2\right)\\+\sin\theta_2\left[t_{p,1}t_{s,2}\left(1-\cos\theta_1\cos\theta_2\right)-t_{s,1}t_{s,2}\left(\cos\theta_1-\cos\theta_2\right)\right]J_2\left(kr\sin\theta_1\right)J_1\left(kr\sin\theta_2\right)\end{Bmatrix}d\theta_1 d\theta_2$$

. (S87)

Although this coupling term is much smaller than the azimuthal T-spin, it should be regarded as the L-spin (19), which is related to the symmetry of modes and possesses $\mathbb{Z}_4$ topological invariance. Noteworthily, the terms of the second row in the integration can cancelled by considering the symmetry. In addition, it can be deduced that

$$S_r = \left(\frac{1}{2k^2}\nabla\times\mathbf{\Pi}\right)_r = 0. \tag{S88}$$

There is no SAM density in the radial direction.

We compare the components of SAM and the T-spin given by the vorticity of kinetic momentum in Fig. S8. The *x*-component of SAMs [Fig. S8(**a**) and Fig. S8(**f**)] and *y*-component of SAMs [Fig. S8(**b**) and Fig. S8(**g**)] are helicity-independent, while the *z*-component SAMs [Fig. S8(**c**) and Fig. S8(**h**)] are helicity-dependent. There will always be an inverted L-spin exist owing to the geometric phase in the focusing system (10), albeit this inverted component is much small in the focusing circular polarization lights of free space. Remarkably, the *z*-component T-spins [Fig. S8(**d**) and Fig. S8(**i**)] can be antiparallel to the spin vector of incident wave. By subtracting the *z*-component T-spins from *z*-component SAM, the remaining spins are purely parallel to the spin vector of incident wave (without antiparallel component) [Fig. S8(**e**) and Fig. S8(**j**)], which are definitely the L-spins. This phenomenon indicates the T-spin is related to the evolution of geometric phase. The **Fig. S9** exhibit the similar phenomena by changing the numerical aperture (NA) to 0.7.

In conclusion, the former results indicate that there are the extraordinary helicity-dependent T-spin in the *z*-direction and the helicity-independent T-spin in the azimuthal direction in the focusing of circularly polarized lights. Thus, the spin-momentum locking property of the focused field for the circularly polarized lights is also helicity-dependent. Remarkably, these two situations are corresponding to the two spin-momentum locking states in Fig. S3(**a**) and Fig. S3(**c**). Together with the other two states by flipping the propagating direction, the four spin-momentum locking states are consistent with the $\mathbb{Z}_4$ topological invariant of generic EM modes.

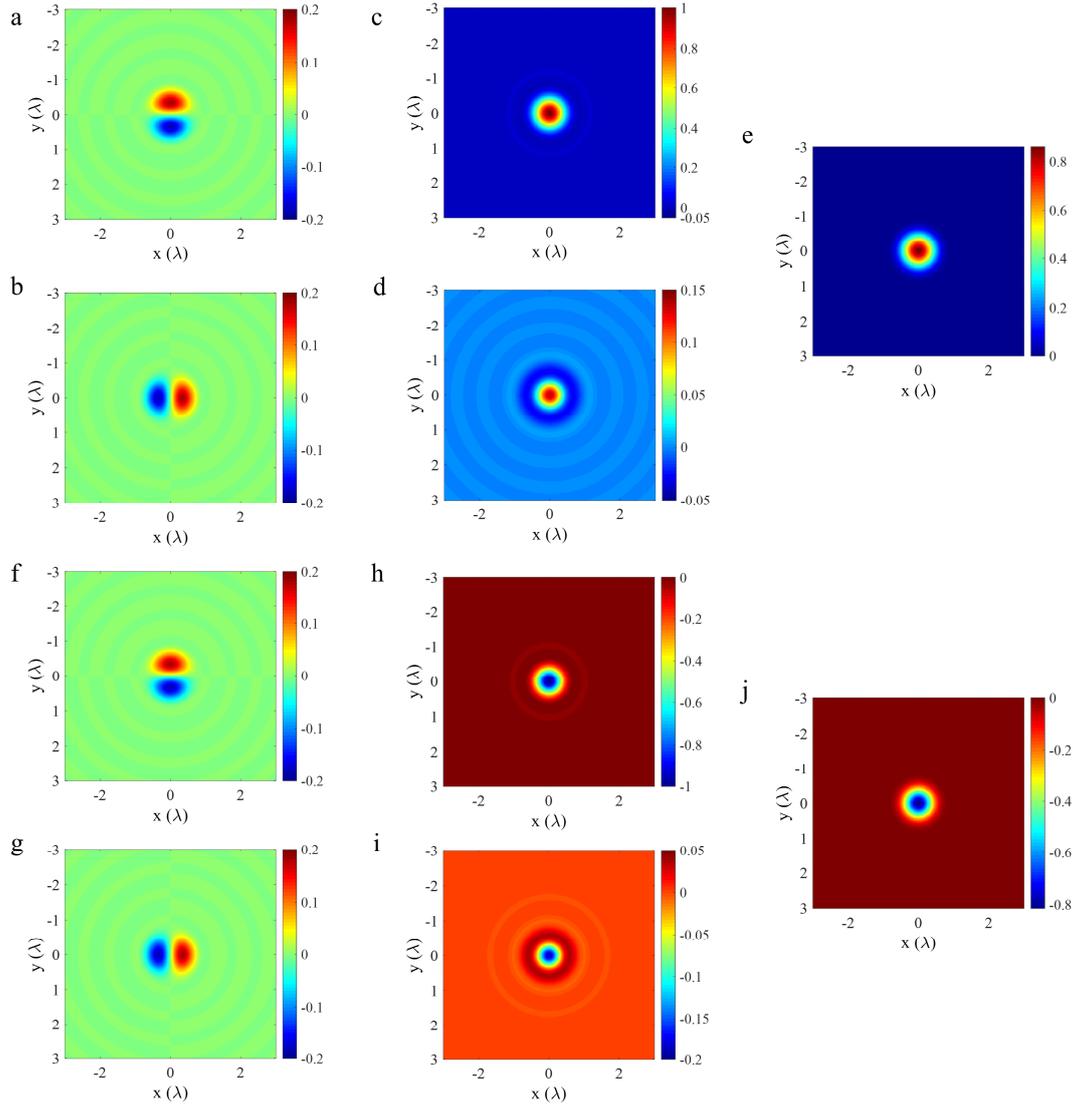

**Fig. S8.** The spin properties of focused circular polarization lights. The (a) $S_x$ and (b) $S_y$, (c) $S_z$, (d) $z$-component of T-spin and (e) L-spin of the focused right-handed circular polarization light. The (f) $S_x$ and (g) $S_y$, (h) $S_z$, (i) $z$-component of T-spin and (j) L-spin of focused left-handed circular polarization light. From S8. (a, b, f, g), it can be found that the T-spins are locked with the propagating direction and helicity-independent. The L-spin is helicity-dependent and inverse as the incident circular polarization is inverted, and the $z$-component T-spins are also helicity-dependent from S8. (d, i). Importantly, the inverted SAM components in L-spins disappear in S8. (e, j) by subtracting the total spins from T-spins, which are related to the geometric phase. The NA is 0.5 and the wavelength is 632.8nm.

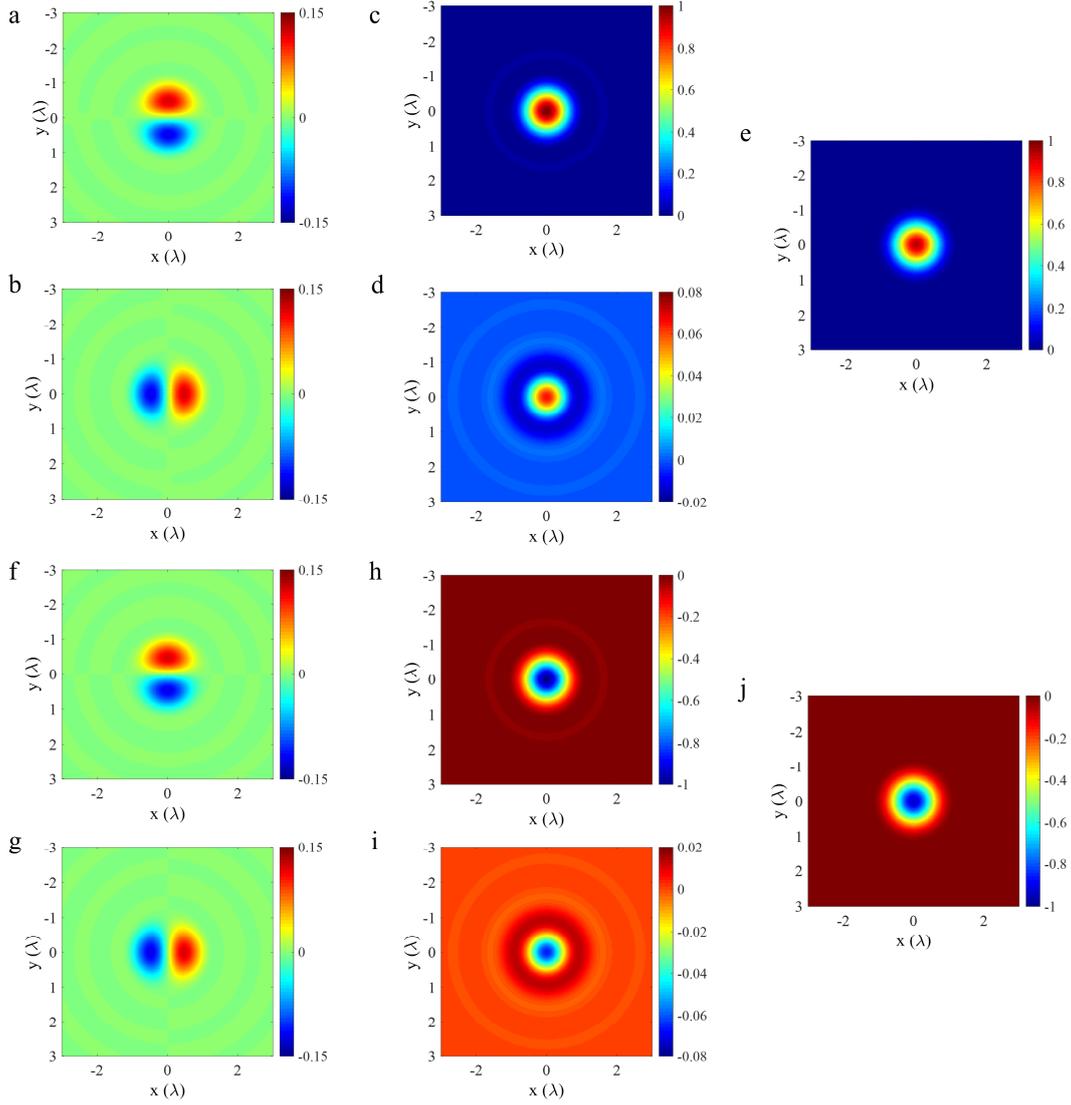

**Fig. S9.** The spin properties of focused circular polarization lights. The (a) $S_x$ and (b) $S_y$, (c) $S_z$, (d) $z$-component of T-spin and (e) L-spin of the focused right-handed circular polarization light. The (f) $S_x$ and (g) $S_y$, (h) $S_z$, (i) $z$-component of T-spin and (j) L-spin of focused left-handed circular polarization light. From S9. (a, b, f, g), it can be found that the T-spins are locked with the propagating direction and helicity-independent. The L-spin is helicity-dependent and inverse as the incident circular polarization is inverted, and the $z$-component T-spins are also helicity-dependent from S9. (d, i). Importantly, the inverted SAM components in L-spins disappear in S9. (e, j) by subtracting the total spins from T-spins, which are related to the geometric phase. Remarkably, these two situations are corresponding to the two spin-momentum locking states in Fig. S3(a) and Fig. S3(c). Together with the other two states by flipping the propagating direction, the four spin-momentum locking states are consistent with the $\mathbb{Z}_4$ topological invariant of generic EM modes. The NA is 0.7 and the wavelength is 632.8nm.

**Supplementary Section 9:** Experimental setup and measurement of SAM perpendicular to optical axis

The experimental setup of the mapping the transverse components of SAM is given in Fig. S10. The incident beam of 632.8nm-wavelength is tightly focused by an objective lens (Olympus, NA=0.5, 50×) to a PS nanoparticle with diameter 201nm sitting on a silver film (thickness: 45nm). The focusing field and the scattering field of the PS particle (the far-field radiation field and part of the near-field evanescent field) radiate downward through the coupling of the silver film. Then the signal is collected by an oil immersion objective lens (Olympus, NA = 1.49, 100×). By a piezo-stage (Physik Instrumente, P-545), we move precisely the PS particle through the focal plane of a tightly focused beam. Each time the position is moved, the back focal plane intensity (far-field intensity) distribution is imaged by a four-quadrant detector. Through the dipole theory (S6) and similar techniques as Ref. (13), the transverse components of SAM can be reconstructed.

The theoretical and reconstructed results for the focused left-handed circularly polarized light and right-handed circularly polarized light can be found in Figs. S11 and S12. These experimental results match well with the theoretical results, which reveals the helicity-independent T-spin in the focusing.

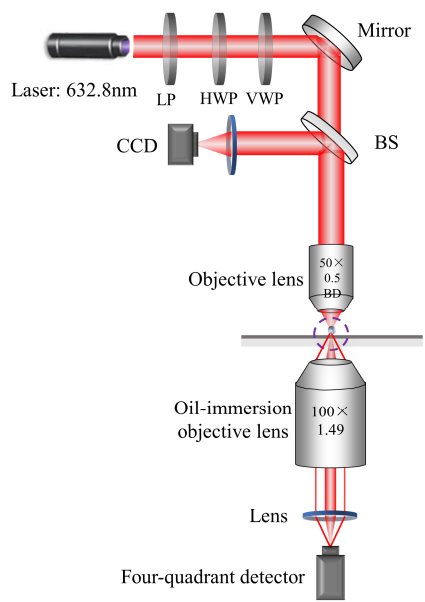

**Fig. S10.** The schematic diagram for mapping the spin component perpendicular to the optical axis. LP: linear polarizer; HWP: Half-wave plate; QWP: quarter-wave plate; BS: unpolarized beam-splitter.

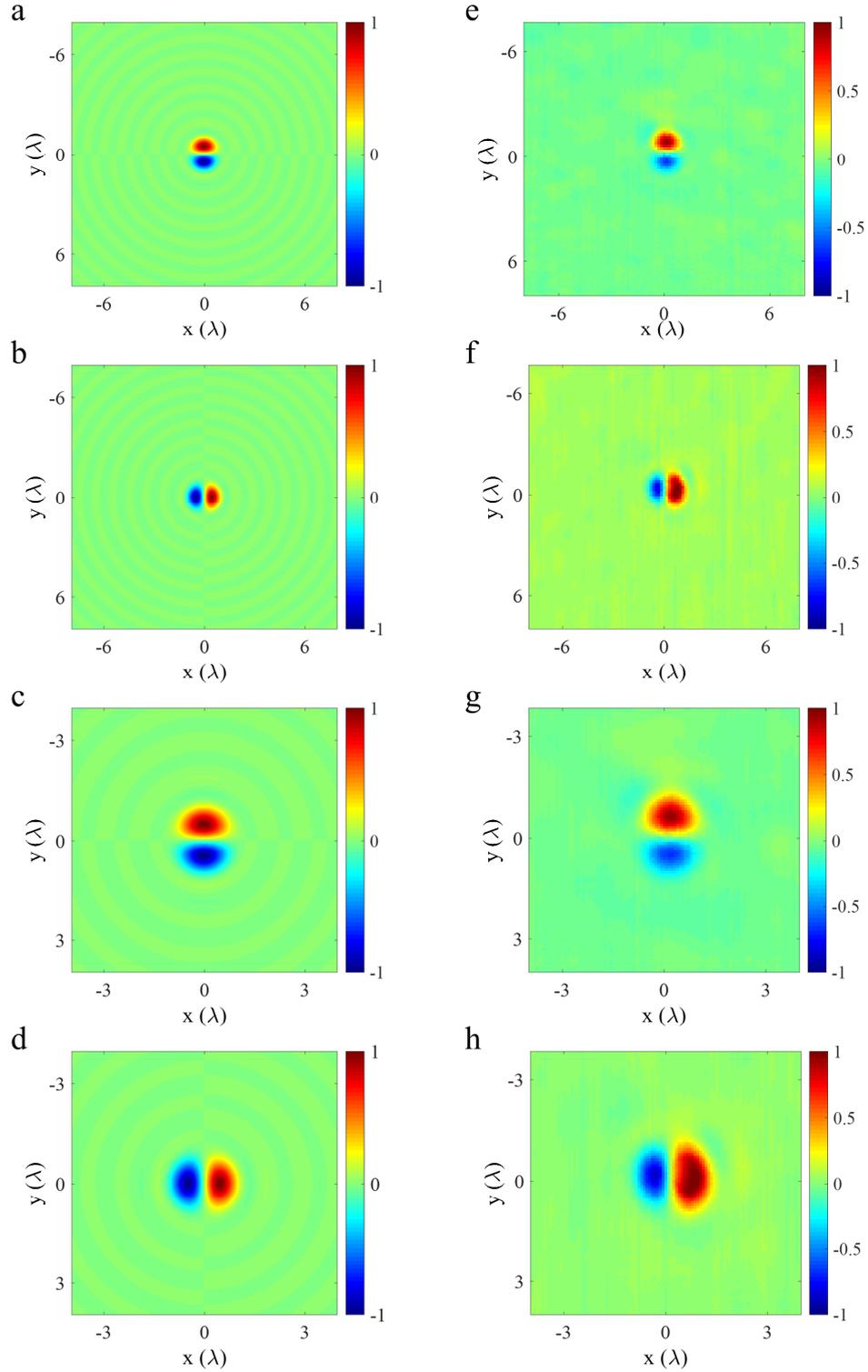

**Fig. S11.** The theoretical (a) $S_x$ and (b) $S_y$, and the experimental (e) $S_x$ and (f) $S_y$ for the focused left-handed circularly polarized light in a region of 10μm×10μm. The grid size is 100nm here. The theoretical (c) $S_x$ and (d) $S_y$, and the experimental (g) $S_x$ and (h) $S_y$ for the focused left-handed circularly polarized light in a region of 5μm×5μm. The grid size is 50nm here. The theoretical results match well with the experimental results. The wavelength is 632.8nm and the numerical aperture is 0.5.

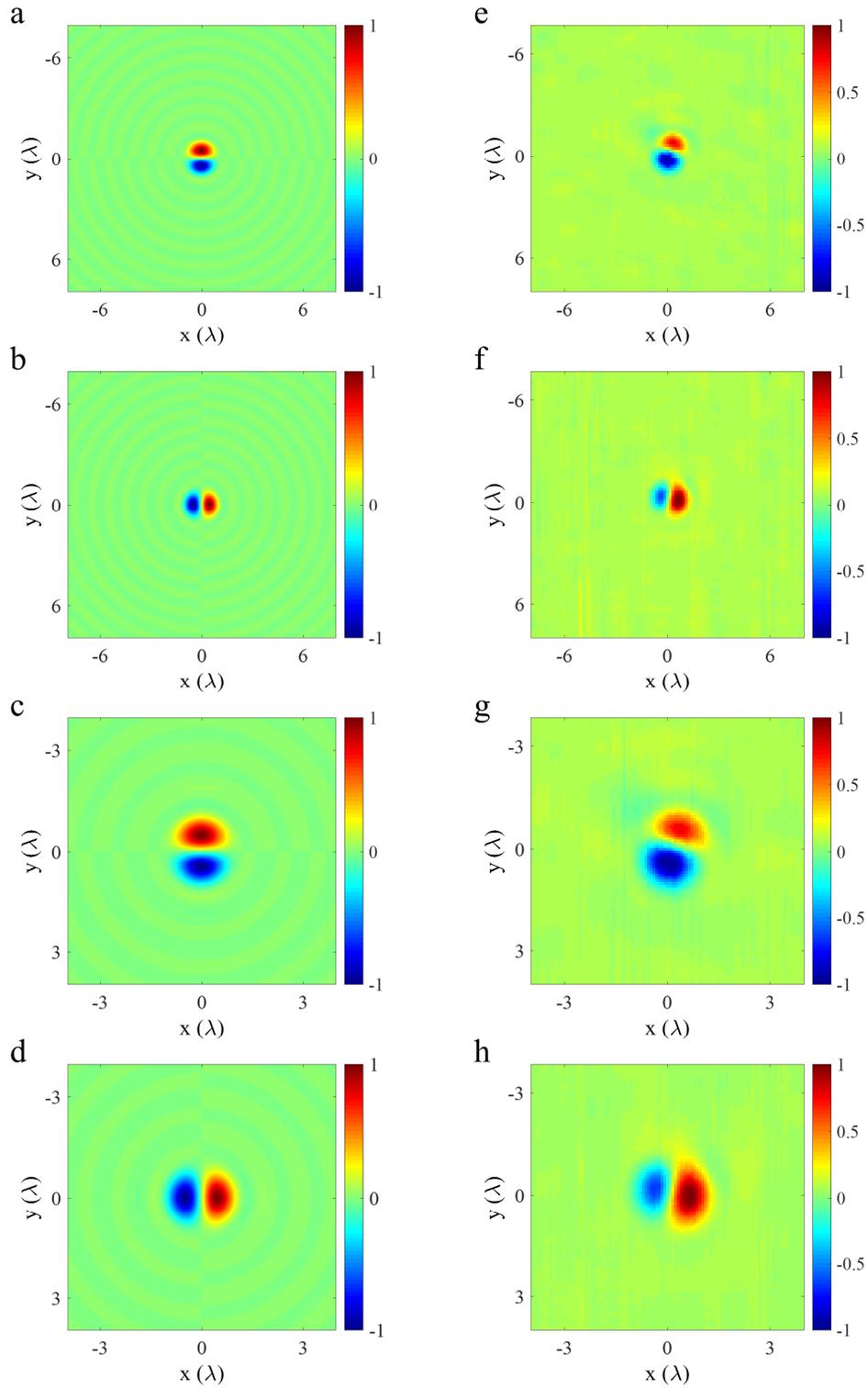

**Fig. S12.** The theoretical (a) $S_x$ and (b) $S_y$, and the experimental (e) $S_x$ and (f) $S_y$ for the focused right-handed circularly polarized light in a region of 10μm×10μm. The grid size is 100nm here. The theoretical (c) $S_x$ and (d) $S_y$, and the experimental (g) $S_x$ and (h) $S_y$ for the focused left-handed circularly polarized light in a region of 5μm×5μm. The grid size is 50nm here. The theoretical results match well with the experimental results. The wavelength is 632.8nm and the numerical aperture is 0.5.

**Supplementary Section 10:** Experimental setup and measurement of SAM parallel to optical axis

The setup of the tip-fiber-based-measurement system to map the out-of-plane component of SAM is shown in Fig. S13(**a**). A He–Ne laser with an operating wavelength of 632.8 nm was used as a light source. The light beam was expanded and collimated via a telescope system. Then it passed through a linear polarizer (LP) and a quarter wave plate (QWP) to produce the desired left-handed or right-handed circularly polarized lights, which was then focused using an objective lens (Olympus, NA 0.7, 60×) onto a silica coverslip properly for the further scanning imaging by a self-assembly near-field scanning optical microscopic (NSOM) system. The NSOM probe with a hole as shown in Fig. S13(**b**) was controlled using a tuning fork feedback system for mapping of the in-plane field distributions of the focused beams. The near-field signal that was coupled through the nanohole to the fiber was split and then analyzed using a combination of a quarter-wave plate and a linear polarizer to extract the individual circular polarization ($I_{LCP}$: left-handed circularly polarized component and $I_{RCP}$: right-handed circularly polarized component) component of the signal. These components were then directed to two photomultiplier tubes (PMTs) to measure the intensity information of the two signals. This then enables characterization of the out-of-plane SAM component (i.e., along the optical axis) of the focused beams by the relation (S7):

$$S_z = \frac{\varepsilon}{4\omega} \frac{k^2 + \kappa^2}{\kappa^2} \left( I_{RCP} - I_{LCP} \right). \tag{S89}$$

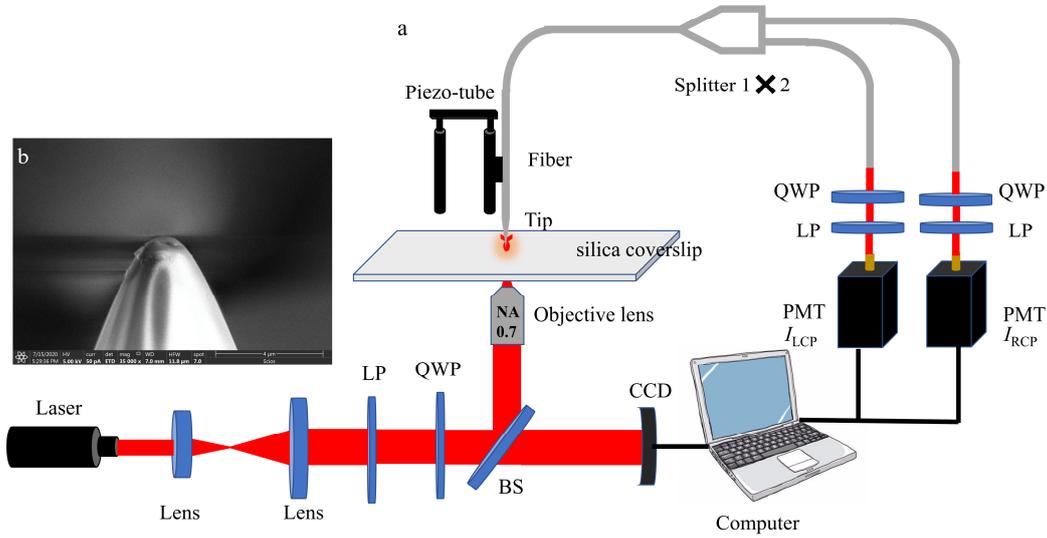

**Fig. S13.** (a) The schematic diagram for mapping the spin component along the optical axis. The details of experimental method can be found in Ref. (S7). LP: linear polarizer; QWP: quarter-wave plate; PMT: photo-multiplier tube; BS: unpolarized beam-splitter. (b) The scanning electron microscope image of optical fiber nanoprobe with a circular hole. The radius of hole is about 50nm.

The measured field components and reconstructed out-of-plane SAM can be found in Fig. S14 and Fig. S15 for the right-handed circularly polarized light and left-handed circularly polarized light, respectively.

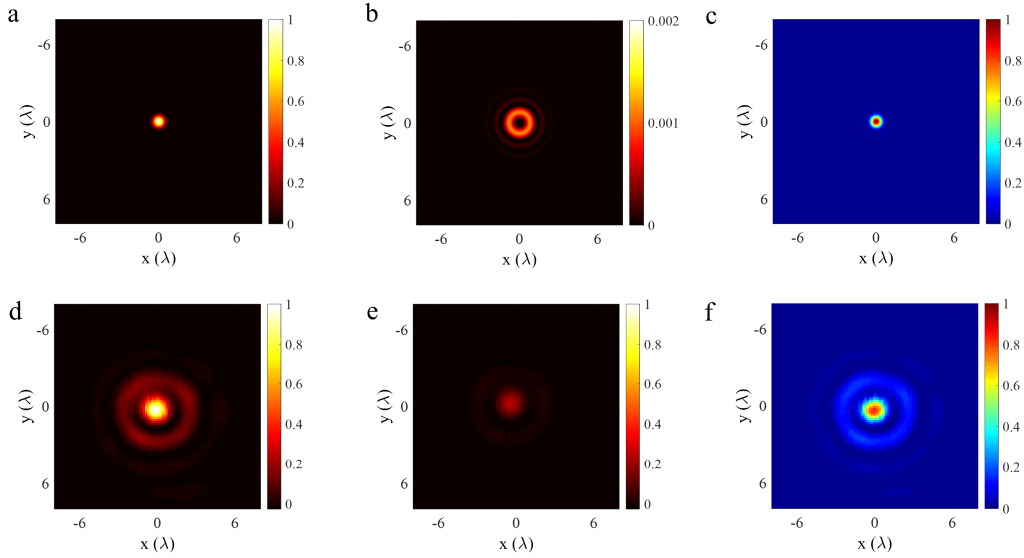

Fig. S14. The theoretical (a) $I_{RCP}$, (b) $I_{LCP}$ (c) normal SAM and measured (d) $I_{RCP}$, (e) $I_{LCP}$ and (f) normal SAM for the focused right-handed circularly polarized light. The numerical apertures of theoretical and experimental systems are 0.7. The dimension of a pixel is 100nm and the scanned region is 10μm×10μm. Note here that the $I_{LCP}$ is much smaller than the $I_{RCP}$.

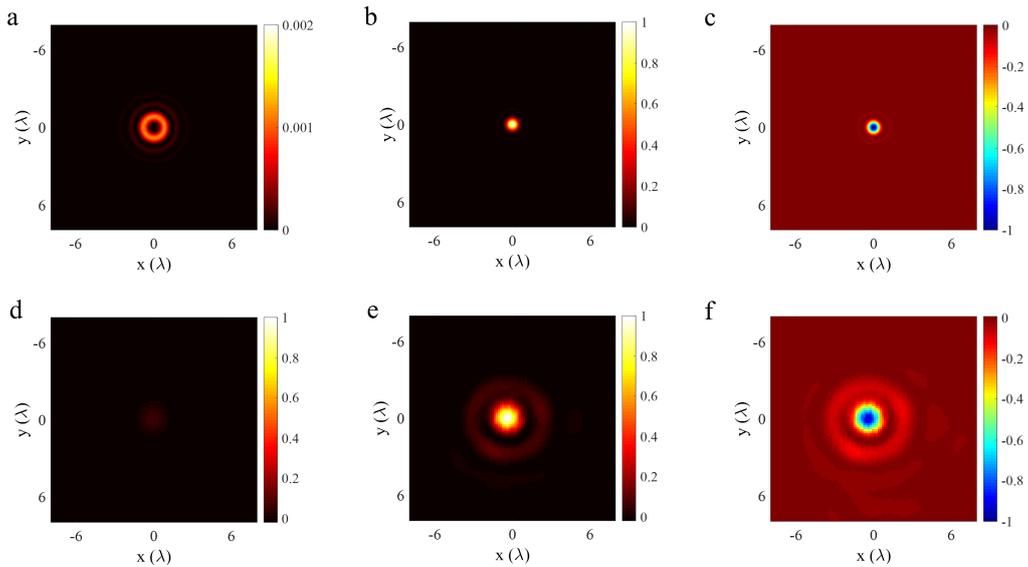

**Fig. S15.** The theoretical (a) $I_{RCP}$, (b) $I_{LCP}$ (c) normal SAM and measured (d) $I_{RCP}$, (e) $I_{LCP}$ and (f) normal SAM for the focused left-handed circularly polarized light. The numerical apertures of theoretical and experimental systems are 0.7. The dimension of a pixel is 100nm and the scanned region is 10μm×10μm. Note here that the $I_{RCP}$ is much smaller than the $I_{LCP}$.

The experimental results match well with the theoretical results, indicating that the L-spin of focused field is inverse as the polarization of incident light switches from the right-handed circular polarization to the left-handed circular polarization. Because there are longitudinal parts of 'T-spins' existing in these focused fields, these results can be regarded as the evidences that the T-spin is helicity-dependent. The distinctions between the theoretical results and measured results are owing to the three reasons: first, the radius of air-hole is about 50nm, and thus each pixel would collect the light at a region of 100nm×100nm.

This will downgrade the contrast of image and raise up the intensity of sidelobes; second, although the aperture-type NSOM will collect the transverse components of optical field primarily. However, the longitudinal component can also pass through the air-hole and affect the distributions of measured results; third, it can hardly put the nanoprobe at the focal plane exactly, and thus the scale of focal point at an out-of-focus plane would be larger compared to that of focal plane.

**Supplementary Section 11:** Transverse optical forces by spin-momentum locked photonic skyrmions

In the former analysis, one can realize that the SAMs of the focused circularly polarized lights are similar to the distributions of Bloch-type magnetic skyrmion (S8). The one-dimensional contours of $S_z$ (red line), $S_\varphi$ (blue line) and $S_r$ (green dotted line) and vectorgraphs for the focused RCP light and focused LCP light can be found in Fig. S16. The spin vectors vary from 'up' state to 'down' state for focused RCP light, while the tendency is reversal as the circular polarization of incident wave is inverted.

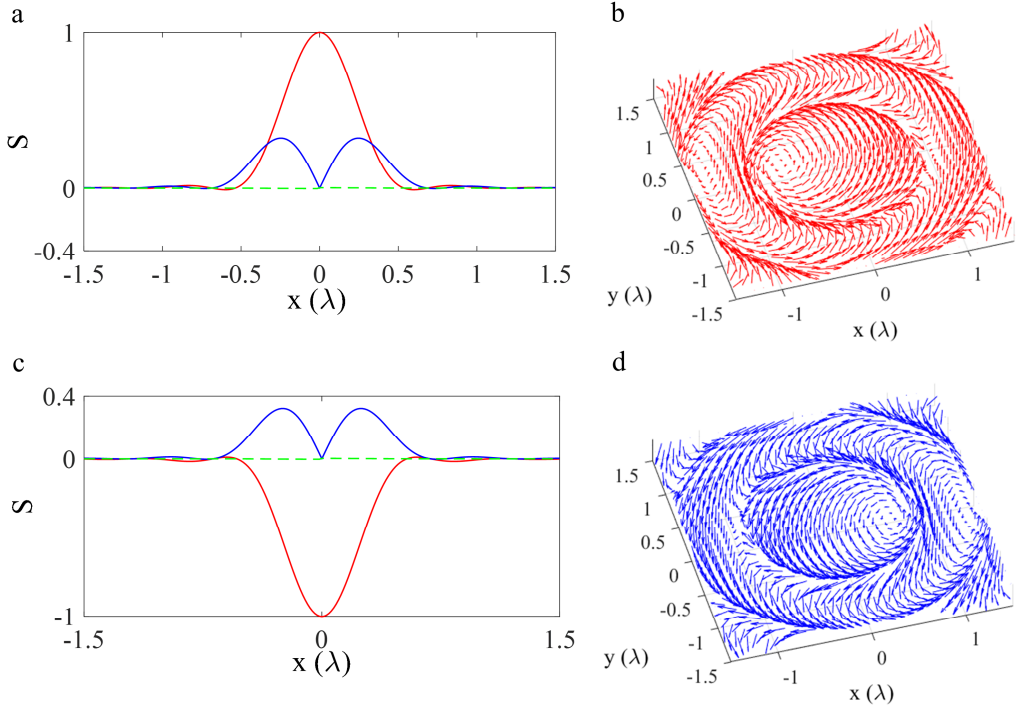

**Fig. S16.** The spin texture of focused circular polarization lights. (a) The one-dimensional contours of $S_z$ (red line), $S_\varphi$ (blue line) and $S_r$ (green dotted line) and (b) vectorgraphs for the focused RCP light; (c) the one-dimensional contours of $S_z$ (red line), $S_\varphi$ (blue line) and $S_r$ (green dotted line) and (d) vectorgraphs for the focused LCP light. The SAMs of the focused circularly polarized lights are similar to the distributions of Bloch-type skyrmion.

In the following, we will give an example for the application of optical spin-momentum locking in free space: the spin-momentum locked transverse optical force for the wideband chiral sorting. The schematic diagram can be found in Fig. S17. The Bloch-type skyrmion with skyrmion number −1 (generated by focused RCP light as shown in Fig. S17(**a**)) can produce the negative transverse optical force $F_x$ as shown in Fig. S17(**b**), while the Bloch-type skyrmion with skyrmion number +1 (generated by focused LCP light as shown in Fig. S17(**f**)) can produce positive transverse optical force $F_x$ as shown in Fig. S17(**g**). The momentum locked wideband unidirectional optical force can be definitely found in the wavelength region (~0.8μm–1.2μm) in the green dotted boxes.

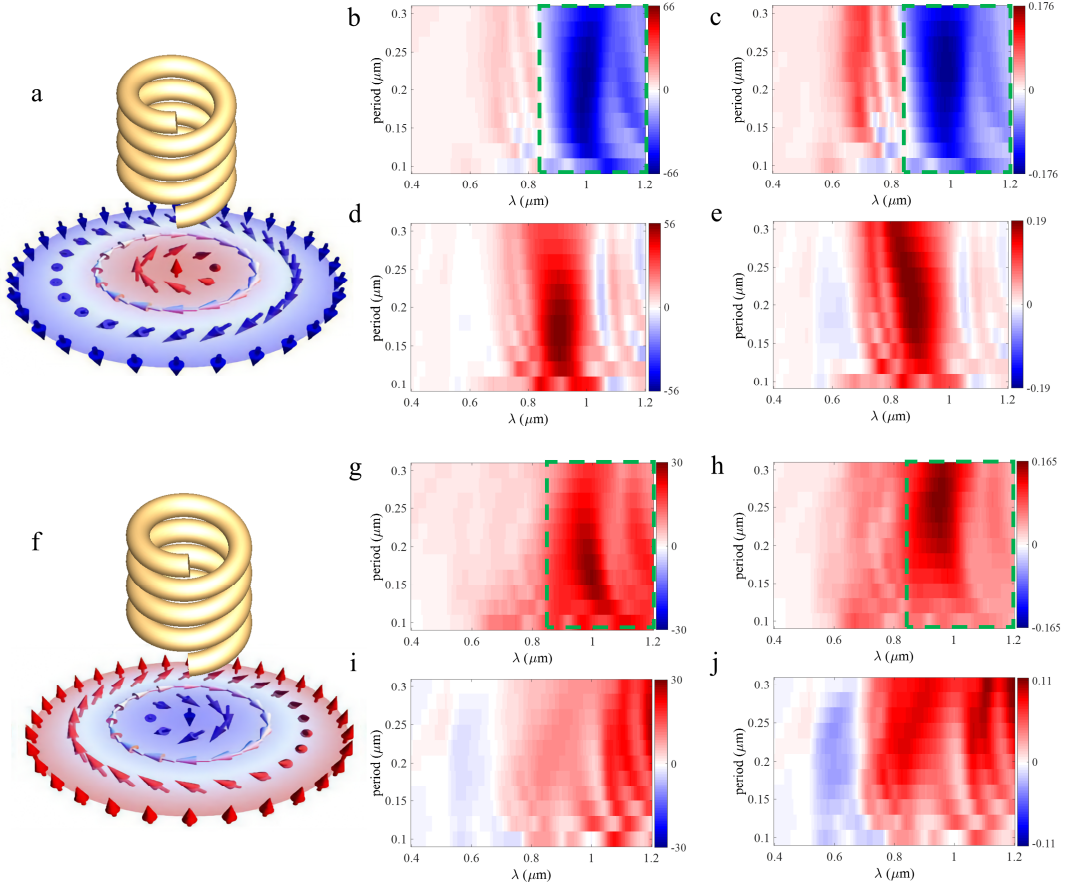

**Fig. S17.** (a) The schematic diagram of −1-order optical Bloch-type skyrmion interacting with the gold helix. (b) The transverse optical force $F_x$ and (c) the ratio between the $F_x$ and longitudinal optical force $F_z$; (d) The transverse optical force $F_y$ and (e) the ratio between the $F_y$ and longitudinal optical force $F_z$. (f) The schematic diagram of +1-order optical Bloch-type skyrmion interacting with the gold helix. (g) The transverse optical force $F_x$ and (h) the ratio between the $F_x$ and longitudinal optical force $F_z$; (i) The transverse optical force $F_y$ and (j) the ratio between the $F_y$ and longitudinal optical force $F_z$. The units for S17. (b, d, g, i) are $10^{-5}$pN·mW$^{-1}$·μm$^2$. The radius of helix in $xy$-plane is 0.2μm; the radius of metal fiber in helix is 0.04μm; and 4 periods are used here.